\documentclass[11pt]{article}
\pdfoutput=1

\usepackage[utf8]{inputenc}
\usepackage{multirow}
\usepackage{amsmath, amsfonts, amssymb}
\usepackage{comment}
\usepackage{graphicx}
\usepackage{float}
\usepackage{psfrag}
\usepackage{amsthm}
\usepackage{caption}
\usepackage{subcaption}
\usepackage{import}
\usepackage{bm}
\usepackage[usenames,dvipsnames,svgnames,table]{xcolor}
\usepackage{enumerate}
\usepackage{arydshln}
\usepackage{soul}
 \usepackage{slashed}
\usepackage{mathrsfs}
\usepackage{mathalfa}
\usepackage{longtable}
\allowdisplaybreaks
 \usepackage{a4wide}
  \usepackage{tikz}
  \usepackage{tikz-cd}
  \usetikzlibrary{shapes.geometric,arrows}
  \usepackage{tcolorbox}
\tikzstyle{startstop} = [rectangle, rounded corners, minimum width=3cm, minimum height=1cm,text centered, draw=black, fill=red!30]
\tikzstyle{io} = [trapezium, trapezium left angle=70, trapezium right angle=110, minimum width=3cm, minimum height=1cm, text centered, draw=black, fill=blue!30]
\tikzstyle{process} = [rectangle, minimum width=3cm, minimum height=1cm, text centered, draw=black, fill=orange!30]
\tikzstyle{decision} = [diamond, minimum width=3cm, minimum height=1cm, text centered, draw=black, fill=green!30]
\tikzstyle{arrow} = [thick,->,>=stealth]
\tikzcdset{row sep/normal=0.75cm}

\usepackage[labelformat=simple]{subcaption}    
  \usepackage{color}
  \definecolor{dark-gray}{gray}{0.20}
  \definecolor{gray}{gray}{0.30}
  \definecolor{light-gray}{gray}{0.80}
  \definecolor{dark-red}{rgb}{0.7,0,0}
  \definecolor{dark-green}{rgb}{0,0.5,0}
  \definecolor{dark-blue}{rgb}{0.3,0.3,0.7}
  \definecolor{light-blue}{rgb}{0.8,0.8,1}
  \definecolor{our-blue}{rgb}{0.118,0.533,0.898}
  \definecolor{our-magenta}{rgb}{0.847,0.106,0.376}
      \definecolor{swamp}{RGB}{240, 199, 197}

  \usepackage{pifont}

\usepackage{newunicodechar} 

\usepackage{setspace}

\newcommand{\be}{\begin{equation}}
\newcommand{\ee}{\end{equation}}

\def\be{\begin{equation}}
\def\ee{\end{equation}}
\def\bea{\begin{eqnarray}}
\def\eea{\end{eqnarray}}

\newcommand{\beq}{\begin{equation}}  \newcommand{\eeq}{\end{equation}}
\newcommand{\bal}{\begin{aligned}}   \newcommand{\eal}{\end{aligned}}
\def\beqa{\begin{eqnarray}}
\def\eeqa{\end{eqnarray}}

\newcommand{\dd}{\mathrm{d}}

\usepackage{stackengine}
\usepackage{calc}
\newlength\shlength
\newcommand\vv[2][0]{\setlength\shlength{#1pt}%
  \stackengine{-5.6pt}{$#2$}{\smash{$\kern\shlength%
    \stackengine{7.55pt}{$\mathchar"017E$}%
      {\rule{\widthof{$#2$}}{.57pt}\kern.4pt}{O}{r}{F}{F}{L}\kern-\shlength$}}%
      {O}{c}{F}{T}{S}}

\def\simleq{\; \raise0.3ex\hbox{$<$\kern-0.75em
      \raise-1.1ex\hbox{$\sim$}}\; }
   \def\simgeq{\; \raise0.3ex\hbox{$>$\kern-0.75em
      \raise-1.1ex\hbox{$\sim$}}\; }

\numberwithin{equation}{section}

\usepackage{jheppub}
\usepackage{hyperref}
\usepackage{cleveref}

\hypersetup{
	colorlinks=true,
	linkcolor=dark-blue,
	citecolor=dark-red,
	urlcolor=dark-green,
	linktoc=page
}

\theoremstyle{remark}

\crefname{appendix}{Appendix}{Appendices}

\title{\centering Cosmology of light towers and swampland constraints }

\author{Gonzalo F. Casas$^1$,} \author{Ignacio Ruiz$^{1,\,2}$} 

\affiliation{$^1$Instituto de F\'{i}sica Te\'{o}rica UAM-CSIC, Universidad Aut\'{o}noma de Madrid, Cantoblanco, 28049 Madrid, Spain}
\affiliation{$^2$Departamento de F\'{i}sica Te\'{o}rica, Universidad Aut\'{o}noma de Madrid, Cantoblanco, 28049 Madrid, Spain}

\preprint{IFT-UAM/CSIC-24-129}
\emailAdd{gonzalo.f.casas@csic.es}
\emailAdd{ignacio.ruiz@uam.es}

\abstract{We study the dynamical evolution of FLRW cosmologies in the presence of a tower of scalar light states and a runaway exponential potential. Some of the attractor solutions have problematic behaviours from the EFT point of view, which we use to argue for restrictions on the possible exponential scalings of the potential and tower characteristic mass as we move towards asymptotic regions in moduli space. These serve as further evidence that the tower mass should not decay faster than the potential or the KK scale associated to the homogeneous decompactification of a single compact dimension. We provide support from different top-down compactifications and connect with previous arguments found in the literature. 

}

\begin{document}
\hypersetup{pageanchor=false}
\makeatletter
\let\old@fpheader\@fpheader

\makeatother
\maketitle

\newcommand{\remove}[1]{\textcolor{red}{\sout{#1}}}

\pagenumbering{roman}

\pagenumbering{arabic}
\setcounter{page}{1} 
\section{Introduction}
\label{sec:intro}

Effective field theories are a useful tool for studying low energy phenomena describing our world. This has allowed important advances in both particle physics and cosmology, without the need for a full understanding of the UV completion of the theory. However, requiring that our EFT can be consistently coupled to a theory of quantum gravity (QG), such as string theory, imposes non-trivial restrictions. This is the goal of the Swampland Program\cite{Vafa:2005ui,Brennan:2017rbf,Palti:2019pca,vanBeest:2021lhn,Grana:2021zvf,Harlow:2022ich,Agmon:2022thq}, which aims to sharply define the boundary between those EFT which have a UV completion (the Landscape) and those which do not (the Swampland). 

The Swampland Program is formulated in terms of conjectures motivated by both top-down (universal patterns found in string theory) and bottom-up (black hole, unitarity, and holography) arguments. Some of these conjectures (coincidentally those better understood) concern the asymptotics of the moduli space, $\mathcal{M}$, the space of massless scalars $\{\varphi^i\}_i$ whose vacuum expectation value (v.e.v.) dynamically control the different parameters of our theory. The geometry of $\mathcal{M}$ is given by the moduli space metric $\mathsf{G}_{ij}$, which can be read from the Einstein frame effective action:
\begin{equation}
    S^{(d)}\supseteq \frac{1}{2}\int\dd^dx\sqrt{-g}\left\{\kappa_d^{-2}\mathcal{R}_g-\mathsf{G}_{ij}\partial_\mu\varphi^i\partial^\mu\varphi^j-2 V(\vec\varphi)\right\}\;,
\end{equation}
where $\kappa_d=M_{{\rm Pl,}d}^{-\frac{d-2}{2}}$ is the gravitational coupling. For simplicity, along this work we will consider canonically normalised moduli, i.e. $\mathsf{G}_{ij}=\kappa_d^{-2}\delta_{ij}$ (note that this is always possible for a single modulus, which is then identified with the geodesic distance, though this is not always guaranteed in multimoduli settings). We also allow some scalar potential $V(\vec\varphi)$ acting as a function of the moduli, which as long as it asymptotically decays to zero ``fast enough'' does not significantly affect the asymptotic geometry of $\mathcal{M}$.

Regarding the asymptotic regions of $\mathcal{M}$, the \textbf{Swampland Distance Conjecture (SDC)} \cite{Ooguri:2006in} states that, given a UV complete $d$-dimensional EFT with moduli space $\mathcal{M}$ and an asymptotic geodesic trajectory on it, an infinite tower of light particles must exist, with mass scale evolving as
\begin{equation}
    \frac{m(\Delta\phi)}{M_{{\rm Pl},d}}\sim e^{-\alpha\Delta\phi}\to 0\quad \text{as}\quad \Delta\phi\to\infty
\end{equation}
where $\Delta \phi$ is the geodesic distance travelled and $\alpha>0$ is a $\mathcal{O}(1)$ number.

This conjecture has been tested and verified in many string theory examples (see references in \cite{Brennan:2017rbf,Palti:2019pca,vanBeest:2021lhn,Grana:2021zvf,Agmon:2022thq}), and has also been motivated from bottom-up arguments \cite{Calderon-Infante:2023ler,Ooguri:2024ofs,Aoufia:2024awo}. 
A refinement of the SDC is the \textbf{Emergent String Conjecture (ESC)} \cite{Lee:2019xtm,Lee:2019wij}, which restricts the nature of the light towers to be either a KK tower resulting from the decompactification of $k$ compact dimensions or the oscillator modes from a unique, critical, weakly coupled string.

While the original formulation of the SDC was meant for massless scalars, it has been shown to also hold for potentials that decay fast enough in asymptotic regions in moduli space, such as 4d $\mathcal{N}=1$ descriptions with positive potentials \cite{Lanza:2021udy,Lanza:2020qmt,Klaewer:2020lfg, Cota:2022yjw}.

Regarding the asymptotic expression of scalar potentials found in string compactifications \cite{Grimm:2019ixq,McAllister:2023vgy,VanRiet:2023pnx}, they are expected to be dominated by exponential terms: 

\begin{equation}
    V(\vec\varphi) = \sum_I V_Ie^{-\vec\lambda^I\cdot\vec\varphi}\quad\text{with constant   }{\lambda}^I_i\in\mathbb{R}\;,
\end{equation}
where again $\{\varphi^i\}_i$ are canonically normalised. This is also motivated by bottom-up arguments \cite{Castellano:2021mmx,Castellano:2022bvr}, and is certainly observed in all known string examples. The fact that this universal shape features no minima is used as motivation against the existence of de Sitter solutions at parametric control in string theory. This is captured by the \textbf{(Asymptotic) de-Sitter Conjecture (dSC)} \cite{Obied:2018sgi,Agrawal:2018rcg} which argues that scalar potentials cannot be arbitrarily flat in asymptotic regions in moduli space: 
\begin{equation}\label{eq. exp pot}
    \frac{\|\vec\nabla V\|}{V}\geq c_d\quad \text{in asymptotic regions of }\mathcal{M}\;,
\end{equation}
where the derivatives and norm are taken with respect to the moduli $\{\varphi^i\}_i$ and $c_d>0$ is some $\mathcal{O}(1)$ number depending only on the spacetime dimension. Note that for a single term in \eqref{eq. exp pot} and a single scalar, $\frac{\|\nabla V\|}{V}=\lambda$. The conjecture has been extensively tested \cite{Maldacena:2000mw,Hertzberg:2007wc,Andriot:2019wrs,Andriot:2020lea,Calderon-Infante:2022nxb,Shiu:2023fhb,Shiu:2023nph,Cremonini:2023suw,Hebecker:2023qke,VanRiet:2023cca,Seo:2024fki,Shiu:2024sbe} and is expected to hold along any asymptotic limit.

As argued in \cite{Ooguri:2018wrx,Hebecker:2018vxz}, both SDC and dSC are related, as in asymptotic regimes of the moduli space, the presence of a large number of new light states dramatically lowers the scale at which semiclassical Einstein gravity breaks down: the species scale \cite{Arkani-Hamed:2005zuc,Distler:2005hi,Dimopoulos:2005ac,Dvali:2007hz,Dvali:2007wp,Dvali:2010vm} given by $\Lambda_{\rm QG}\sim M_{{\rm Pl}, d} N^{-\frac{1}{d-2}}$, with $N$ the number of light states below $\Lambda_{\rm QG}$. As the number $N$ grows exponentially from the light tower as $\Delta\phi\to \infty$, then $\Lambda_{\rm QG}\sim M_{{\rm Pl}, d} e^{-\mu\Delta\phi}\to 0$  for some $\mu\sim\mathcal{O}(1)$ \cite{vandeHeisteeg:2023ubh,Calderon-Infante:2023ler,Basile:2023blg}. Asking the Hubble scale to be below $\Lambda_{\rm QG}$, one has
\begin{equation}
    \kappa_d\sqrt{V}\lesssim H\lesssim \Lambda_{\rm QG}\sim M_{{\rm Pl}, d} e^{-\mu\Delta\phi}\to 0\;,\label{jerarquia}
\end{equation}
and thus it is expected that $V$ also decays exponentially with $\Delta \phi$. See section \ref{sec:dis} for further restrictions $\Lambda_{\rm QG}$ might impose on $(\alpha,\lambda)$.

When first formulated, both conjectures only claimed for the exponential factors $\alpha$ and $\lambda$ to be $\mathcal{O}(1)$. Further refinements proposed lower bounds for them: 
\begin{itemize}
    \item The \textbf{Sharpened Distance Conjecture} \cite{Etheredge:2022opl} argues for $\alpha \geq \frac{1}{\sqrt{d-2}}$ . This is supported by preservation under dimensional reduction and different top-down examples in string/M-theory compactifications. The bound is saturated for emergent string limits, while decompactifications always feature $\alpha>\frac{1}{\sqrt{d-2}}$.
    \item The \textbf{Strong de-Sitter Conjecture}  proposes $c_d=\frac{2}{\sqrt{d-2}}$ \cite{Rudelius:2021azq,Bedroya:2019snp}. Coincidentally, this prevents asymptotic accelerated expansion \cite{Achucarro:2018vey}, and has been checked to hold in all examples in the literature (see above references).
\end{itemize}

The next natural question is about the existence of upper bounds in $\alpha$ and $\lambda$, i.e., whether potentials or towers that decay arbitrarily fast can exist. This is precisely the motivation of this paper. A simple exercise of dimensional reduction (see \cite{Etheredge:2022opl,Castellano:2023jjt,Etheredge:2024tok} for more details) shows that compactifying a $D$-dimensional EFT action on a flat and homogeneous $n$-manifold results in a KK tower scaling as
\begin{equation}\label{e.KK}
    \frac{m_{KK}(\rho)}{M_{\text{Pl},d}}\sim \exp\left(-\sqrt{\frac{d-2+n}{n(d-2)}}\rho\right)\;,\quad \text{where   }\rho=\sqrt{\frac{d-2+n}{n(d-2)}}\log \mathcal{V}_n
\end{equation}
is the canonically normalised radion modulus and $\mathcal{V}_n$ is the volume of the compact $n$-manifold in $M_{{\rm Pl},D}$ units. This on principle sets a naive maximum exponent of $\sqrt{\frac{d-1}{d-2}}$ by taking $n=1$ \cite{Etheredge:2022opl}, and indeed this is observed in all string compactifications. However, unlike the lower bound $\frac{1}{\sqrt{d-2}}\leq \alpha$, there is no clear argument against larger values.\footnote{In \cite{vandeHeisteeg:2023ubh,Calderon-Infante:2023ler} upper bounds for the asymptotic exponential rate $\mu$ of the Specie scale \eqref{jerarquia} were studied, but this does not immediately translate to upper bounds on $\alpha$ or $\lambda$, as tower and potential scales fall faster or at the same order as $\Lambda_{\rm QG}$.} Most of the compactifications studied in the literature feature flat compact manifolds and decompactification limits to empty, flat vacua. Only recently \cite{Etheredge:2023odp, Alvarez-Garcia:2023qqj} more involved decompactification limits have started to be found and studied. In \cite{Etheredge:2023odp}, an explicit expression for $\alpha$ not following \eqref{e.KK} was found, though it still fulfils the upper $\sqrt{\frac{d-1}{d-2}}$ bound. It is not clear, however, that all decompactification limits, including warped ones, must obey this.

The goal of this paper is to further \emph{sharpen} the possible values that $(\alpha,\lambda)$ can take and argue for an upper bound based on a bottom-up approach. To achieve this, we will study the impact that the asymptotic tower of states has on the cosmological evolution. When working well below the quantum gravity cut-off $\Lambda_{\rm QG}$, the light states of the tower contribute to the energy density of the Universe.\footnote{See \cite{Kolb:1983fm, Dienes:1998vg,Servant:2002aq, Gonzalo:2022jac, Obied:2023clp} and references therein and thereof for considerations of light KK states as dark matter, as well as \cite{Casas:2024xqy} in the context of transient dS solutions.} By studying the cosmological evolution of these towers, important insights can be obtained, learning \emph{what goes wrong} when towers decay \emph{too fast}. The analysis of cosmological dynamics in the presence of moduli has been extensively studied in the literature \cite{Copeland:1997et,Rudelius:2022gbz,Shiu:2023fhb}, also including more realistic ingredients, such as matter, radiation or spatial curvature \cite{Marconnet:2022fmx,Andriot:2023wvg,Andriot:2024jsh,Tonioni:2024huw}. 

\begin{figure}
    \centering
    \includegraphics[width=0.65\linewidth]{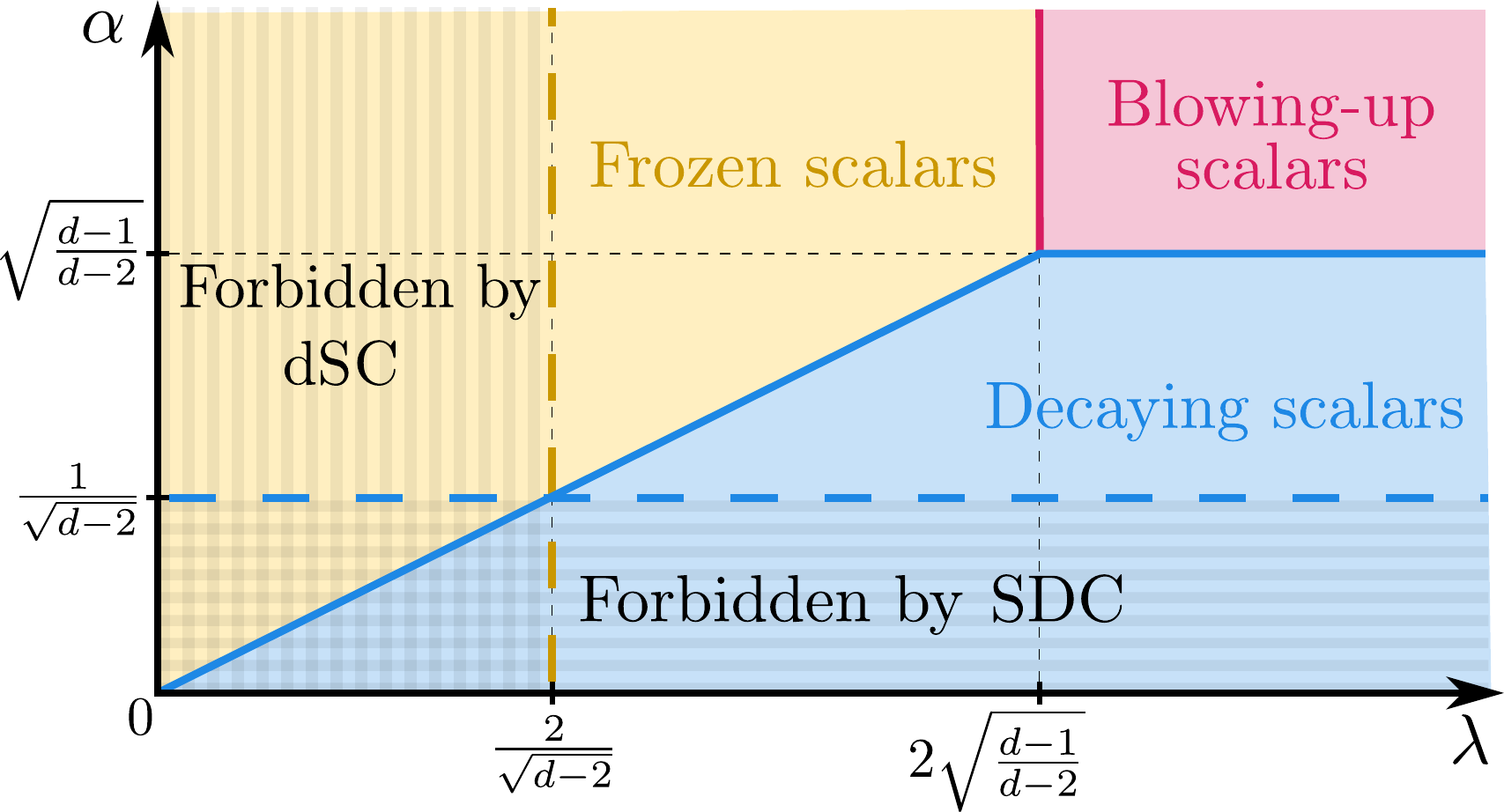}
    \caption{Different regimes regarding the behaviour of the scalar states in the tower as a function of the exponential rates $\lambda$ (scalar potential) and $\alpha$ (light tower). Parameter space regions excluded by Swampland conjectures are also marked.}
    \label{f.exp_space}
\end{figure}

However, in our approach, we will be interested in the evolution of the tower of states (see also \cite{Rudelius:2022gbz}) and the scaling of the potential relative to the scale of the tower. We characterise the different cosmological solutions in terms of the exponential rates of both, $(\lambda, \alpha)$ and their asymptotic signatures at late times, see figure \ref{f.exp_space}. From this, we argue why only some regions of the parameter space $(\lambda, \alpha)$ might be consistent with a UV complete EFT description, and we collect various examples in the string theory literature to strengthen this point. The classification of the allowed possibilities for potentials and towers was already studied in the context of asymptotic Hodge theory \cite{Grimm:2019ixq}, maximal supergravity compactifications \cite{Etheredge:2022opl}, as well as concerning the species scale \cite{Calderon-Infante:2023ler}.

The structure of the paper is as follows. In section \ref{sec:sols} we study the cosmological dynamics of a single modulus in the presence of a tower of state plus a run away potential. In section \ref{sec:dis} we discuss the results and argue for bottom-up constraints in the ($\alpha,\lambda$)-space, as well as their physical implications. In section \ref{sec:pop} we assemble different $(\alpha,\lambda)$ values found in the literature and confront them with the previously obtained constraints. Finally in section \ref{sec:conc} we conclude by summarising our results and highlighting possible future directions.

\section{Cosmological evolution of light towers}\label{sec:sols}
To start our analysis, we consider an $d$-dimensional effective action in Einstein frame with a single (canonically normalised) scalar $\phi$, which can be subject to a scalar potential $V(\phi)$:
\begin{equation}\label{e. EFT action}
    S^{(d)}\supseteq \frac{1}{2}\int \dd^d x\sqrt{-g}\left\{\kappa_d^{-2}\left[\mathcal{R}_g-(\partial\phi)^2\right]-2 V(\phi)-m_0^2 e^{-2\alpha\phi}\sum_n f(n)^2 \psi_n^2-\sum_n\left(\partial\psi_n\right)^2\right\}\;.
\end{equation}
We have further considered a \textcolor{black}{bosonic} tower of states\footnote{Along this paper we will not explicitly assume the number of states of the (possibly infinite) tower, as this will have no impact on the results, as long as all of them scale in the same way with respect to the scalars.} $\{\psi_n\}_n$ with masses parameterised by $\phi$, $\{m_n=m_0 f(n) e^{-\alpha \phi}\}$, where $f(n)$ is some monotonously growing function. The degeneracy of each level of the tower is simply given by the number of states in the tower with the same mass, i.e., $d_n=\#f^{-1}(f(n))$. We are considering a mass and degeneracy parametrization as general as possible, as it will not be relevant to our initial analysis. For KK towers decompactifying $k$ dimensions and obeying Weyl's law\footnote{See \cite{DeLuca:2024fbc} for general comments on Weyl's law for non-smooth compact manifolds.} one would have $f(n)\approx n$ and $d_n\sim n^{k-1}$, while for string oscillator modes $f(n)\sim \sqrt{n}$ and $d_n\sim n^{-\frac{1}{2}(D_{\rm crit}+1)}e^{a\sqrt{n}}$ for some $a>0$ constant and $D_{\rm crit}$ the critical dimension of the string \cite{Green:1987sp}.  Due to the large degeneracy of oscillator modes, the species scale is of the same scale as the tower, $\Lambda_{\rm QG}\sim m_{\rm osc}$, and thus this kind of limits cannot feature states of the tower parametrically below the QG cut-off \textcolor{black}{$\Lambda_{\rm QG}$ \eqref{jerarquia}, which will be taken to be that of the above effective action}. Following the ESC, the limits of interest for this paper will consider in decompactification ones.

As argued in section \ref{sec:intro}, we will study the dynamics at asymptotic regimes of moduli space, $\phi\to \infty$, such that the potential is an exponential runaway, $V(\phi)=V_0e^{-\lambda\phi}$. \textcolor{black}{For simplicity, we will not consider fermionic states in the dynamics since we expect similar results as in the bosonic case. Cosmological solutions with fermions are considered for instance in fermionic-tensor-theories, see \cite{Saha:2019ztr,deSouza:2008az} for more details}. Finally, \textcolor{black}{assuming that the compact dimensions are initially \emph{small enough} compared with the macroscopic (Hubble) scale,} we will take a cosmological ansatz consisting of a $d$-dimensional and spatially flat FLRW background, with spatially homogeneous fields and metric
\begin{equation}
    \dd s^2=-\dd t^2+a(t)^2\left(\dd r^2+r^2\dd \Omega_{d-2}^2\right)\;.
\end{equation}
The Hubble factor is defined as $H=\frac{\dot a}{a}$, where $\dot a\equiv\partial_t a$. The equations of motion are then
\begin{subequations}
    \begin{align}
        \ddot\phi+(d-1)H\dot\phi-\lambda \kappa_d^2V_0 e^{-\lambda\phi}-\alpha m_0^2 e^{-2\alpha\phi}\sum_n f(n)^2\psi_n^2&=0\label{e:phi}\\
        \ddot\psi_n+(d-1)H\dot \psi_n+m_0^2 f(n)^2e^{-2\alpha\phi}\psi_n&=0\label{e:psin}\\
        (d-1)(d-2)H^2-\left[\dot\phi^2+\kappa_d^2\left(\sum_n\dot\psi_n^2+2V_0 e^{-\lambda\phi}+m_0^2e^{-2\alpha\phi}\sum_n f(n)^2\psi_n^2\right)\right]&=0\label{e:F1}\\
        2(d-2)\dot H+(d-1)(d-2)H^2+\left[\dot\phi^2+\kappa_d^2\left(\sum_n\dot\psi_n^2-2V_0 e^{-\lambda\phi}-m_0^2e^{-2\alpha\phi}\sum_n f(n)^2\psi_n^2\right)\right]&=0 \label{e:F2}
    \end{align}
\end{subequations}
The case with no scalar potential corresponds to $V_0=0$. The asymptotic properties of our system can be easily studied by introducing the following kinematic variables \cite{Copeland:1997et},
\begin{equation}\label{e.kin var}
    \begin{array}{ll}
       x=\dfrac{\dot\phi}{\sqrt{(d-1)(d-2)}H}\,,  &  y=\dfrac{\kappa_d\sqrt{\sum_n\dot\psi_n^2}}{\sqrt{(d-1)(d-2)}H}\,,\\
         z=\sqrt{\dfrac{2V_0}{(d-1)(d-2)}}\dfrac{\kappa_de^{-\frac{\lambda}{2}\phi}}{H}\,, & u=\dfrac{\kappa_d m_0 e^{-\alpha\phi}\sqrt{\sum_nf(n)^2\psi_n^2}}{\sqrt{(d-1)(d-2)}H}\;,
    \end{array}
\end{equation}
with phase space $\Gamma=\{\Vec{X}=(x,y,z,u)\in \mathbb{S}^4: -1\leq x\leq 1,\,0\leq y,\,z,\,u\leq 1\}$. These variables correspond to the different contributions, in Hubble units, to the overall cosmological energy budget at some given time $t$. Note that the restriction $x^2+y^2+z^2+u^2=1$ is nothing but the first Friedmann equation \eqref{e:F1}. This allows us to write the following dynamical system:
\begin{subequations}\label{eqs.dyn sis}
    \begin{align}
        \frac{\dd x}{\dd N}&=\frac{d-1}{2}x(x^2+y^2-z^2-u^2-1)+\sqrt{(d-1)(d-2)}\left(\frac{\lambda}{2}z^2+\alpha u^2\right)\\
        y\frac{\dd y}{\dd N}+u\frac{\dd u}{\dd N}&=\frac{d-1}{2}y^2(x^2+y^2-z^2-u^2-1)\notag\\&\qquad +u^2\left[\frac{d-1}{2}(x^2+y^2-z^2-u^2+1)-\alpha\sqrt{(d-1)(d-2)}x\right]\\
        \frac{\dd z}{\dd N}&=\frac{z}{2}\left[(d-1)(x^2+y^2-z^2-u^2+1)-\lambda\sqrt{(d-1)(d-2)}x \right]\;,
    \end{align}
\end{subequations}
where $N=\log a$ denotes the number of $e$-folds travelled. Denoting respectively the energy density fraction given by the kinetic and mass of terms of the tower, the $y$ and $z$ variables are not independent and cannot be separately studied. A thorough analysis shows the only critical loci are those detailed in Table \ref{tab:fixed}, where the existence and stability requirements are also given.\footnote{Because of the coupling between $y$ and $z$, it is not possible to rewrite \eqref{eqs.dyn sis} as $\frac{\dd \vec X}{\dd N}=\vec F(\vec X)$ in such a way that the stability of the critical points $\vec X_0$, with $\vec F(\vec X_0)=\vec 0$, can be studied from the eigenvalues of $\vec\nabla \vec F|_{\vec X_0}$. However, as all critical points have $u=0$, we can restrict to the surface $\Gamma\cap \{u=0\}$ and study stability there.}

\begin{table}[ht]
    \centering
    
   \resizebox{\textwidth}{!}{\begin{tabular}{|p{40mm}|c|c|c|}
    \hline
    \rowcolor{gray!10!}
       \centering Name  & $(x_0,y_0,z_0,u_0)$ & Existence & Stability\\
       \hline
        \centering Tracker $T_\lambda$  & $\left(\frac{\lambda}{2}\sqrt{\frac{d-2}{d-1}},0,\sqrt{1-\frac{\lambda^2}{4}\frac{d-2}{d-1}},0\right)$ & $\lambda\in\left[0, 2\sqrt{\frac{d-1}{d-2}}\right]$  & Stable \\\hline
      \centering  Modulus Kination $K_\pm$&$(\pm 1,0,0,0)$ & $\lambda \geq 2\sqrt{\frac{d-1}{d-2}} $ & $K_+$ Stable, $K_-$ Unstable\\\hline
       \centering  Tower Kination $K_\theta$  & \centering$\left(\cos\theta,\sin\theta,0,0\right)$ & $\lambda \geq 2\sqrt{\frac{d-1}{d-2}} $, $\alpha>\sqrt{\frac{d-1}{d-2}}$  &
       
      $ \cos \theta\in\left(\max\{\alpha^{-1},\frac{2}{\lambda}\}\times\sqrt{\frac{d-1}{d-2}},1\right)$
       \\
         \hline
    \end{tabular}}
    \caption{Fixed points of dynamical system \eqref{eqs.dyn sis}, as well as conditions for existence and stability. Regarding the $K_\theta$ outside the stability region $\cos \theta\geq\max\{\alpha^{-1},\frac{2}{\lambda}\}\times\sqrt{\frac{d-1}   {d-2}}$, it is unstable outside it and a saddle point when saturating the inequality.}
    \label{tab:fixed}
\end{table}

\begin{figure}[h]
\begin{center}
\begin{subfigure}[b]{0.49\textwidth}
\center
\includegraphics[width=0.95\textwidth]{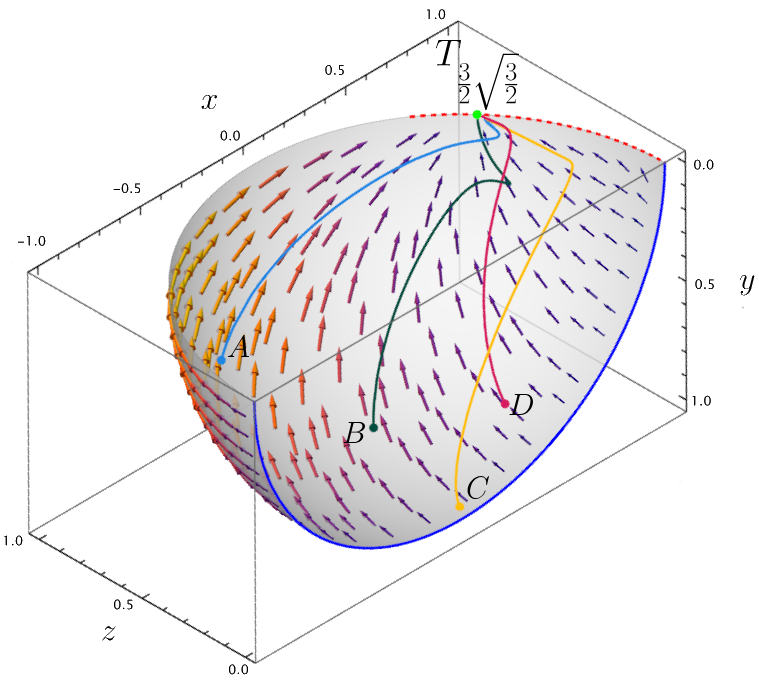}
\caption{\hspace{-0.3em}) $T_\lambda$ with $d=4$, $\lambda=\frac{3}{2}\sqrt{\frac{3}{2}}$ and $\alpha=1$.} \label{f.din1}
\end{subfigure}
\begin{subfigure}[b]{0.49\textwidth}
\center
\includegraphics[width=0.95\textwidth]{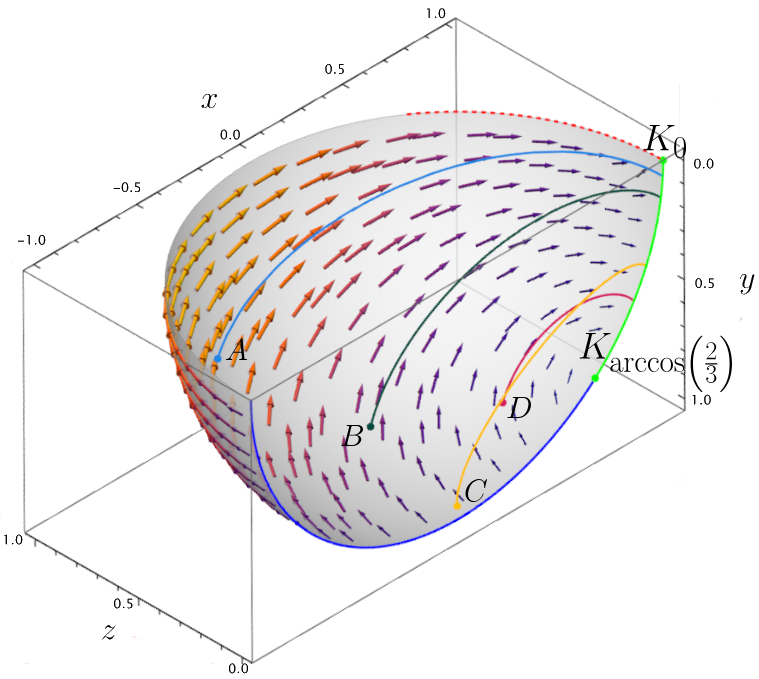}
\caption{\hspace{-0.3em}) $K_\theta$ with $d=4$, $\lambda=3\sqrt{\frac{3}{2}}$ and $\alpha=\frac{3}{2}\sqrt{\frac{3}{2}}$.} \label{f.din2}
\end{subfigure}
\caption{Dynamical system for the $\vec X=(x,y,z,u)\in \Gamma$ variables (note that the radial direction corresponds to $\sqrt{1-u^2}$). The trajectories towards the tracker $T_\lambda$ and tower kination $K_\theta$ attractors (in green) are depicted, with starting points $A=(-0.572, 0.333, 0.589, 0.464)$, $B=(0, 0.822, 0.421,0.384)$, $C=(0.041, 0.989, 0.044,0.135)$ and $D=(0.437, 0.840, 0.218,0.236)$. Note that for the tower kination solutions, the actual $K_{\theta_0}$ attractors depend on the initial conditions. The vector field associated with the dynamical system \eqref{e.kin var} is depicted for the $u=0$ slice.
\label{f.sin}}
\end{center}
\end{figure}

Two attractors and the trajectories towards them are depicted in Figure \ref{f.sin}.
    All critical points lay in the boundary of $\Gamma\cap\{u=0\}$. Note that $K_\theta$ is unstable for all $\theta$ when $\lambda<2\sqrt{\frac{d-1}{d-2}}$, and that for any value of $\lambda$ there is always at least one attractor point, with the transition between tracker and kination solutions occurring precisely at $\lambda=2\sqrt{\frac{d-1}{d-2}}$.

    The second Friedmann equation \eqref{e:F2} can be written as
    \begin{equation}
        \frac{\dot H}{H^2}=-\frac{d-1}{2}(x^2+y^2-z^2-u^2+1)=-(d-1)(x^2+y^2)\;
    \end{equation}
so that for a critical point $\vec X_0$ we have
\begin{subequations}
\begin{align}    
    H&=\frac{H_0}{1+H_0(d-1)(x_0^2+y_0^2)(t-t_0)}\\
    \phi&=\phi_0+\sqrt{\frac{d-2}{d-1}}\frac{x_0}{x_0^2+y_0^2}\log\left[1+H_0(d-1)(x_0^2+y_0^2)(t-t_0)\right]\label{e.at phi}
\end{align}
\end{subequations}
 The fixed points $T_\lambda$ and $K_{0,\pi}$ where already present in the literature \cite{Copeland:1997et}, and correspond to tracker \cite{Wetterich:1987fm,Ferreira:1997hj,Copeland:1997et} and kination \cite{Joyce:1996cp,Gouttenoire:2021jhk} solutions involving a single scalar. Both of them have been studied in the context of the early Universe and string cosmology in general (see \cite{Apers:2022cyl,Cicoli:2023opf,Apers:2024ffe} and references therein) and are well understood. The points $K_\theta$ for $\theta\neq 0,\pi$ are novel and are associated to the tower of states having asymptotically non-vanishing kinetic energy (in Hubble units). 

 While in the classification of the attractor points the value of $\alpha$ has only come into play for $K_\theta$, we will show now that indeed the exponential rate of the tower becomes relevant for the dynamics of the fields $\{\psi_n\}_n$ in the different attractor points. Consider for this the different attractor points:
\begin{itemize}
    \item \textbf{Tracker solution} \cite{Wetterich:1987fm,Ferreira:1997hj,Copeland:1997et}: For $0\leq\lambda\leq 2\sqrt{\frac{d-1}{d-2}}$ the point $T_\lambda$ is an attractor, with
    \begin{subequations}
    \begin{align}        
        H(t)&=\frac{H_0}{1+\frac{d-2}{4}\lambda^2H_0(t-t_0)}\\
        \phi(t)&=\phi_0+\frac{2}{\lambda}\log\left[1+\frac{d-2}{4}\lambda^2H_0(t-t_0)\right]\;.
    \end{align}
    \end{subequations}
    As for \eqref{e:psin}, it can be rewritten as
    \begin{equation}
        \ddot\psi_n+\frac{(d-1)H_0}{1+\frac{d-2}{4}\lambda^2 H_0(t-t_0)}\dot\psi_n+m_0^2 f(n)^2\left[1+\frac{d-2}{4}\lambda^2 H_0(t-t_0)\right]^{-4\frac{\alpha}{\lambda}}\psi_n=0\;,
    \end{equation}
    or equivalently, in terms of the moduli $\phi$,
    \begin{equation}
        \psi_n''+\tilde{\lambda}\psi_n'+\mu_n^2e^{(\lambda-2\alpha)\phi}\psi_n=0\;,
    \end{equation}
where $g'=\partial_\phi g$, $\tilde{\lambda}=\frac{\lambda}{2}\left[\left(\frac{2}{\lambda}\sqrt{\frac{d-1   }{d-2}}\right)^2-1\right]>0$ and $\mu_n=\frac{2m_0 f(n)}{(d-2)\lambda H_0}$. The above equation has as solutions
\begin{equation}
    \psi_n(\phi)=e^{-\frac{\tilde{\lambda}}{2}(\phi-\phi_0)}\left[\mathsf{c}_-J_{-\frac{\Tilde{\lambda}}{\lambda-2\alpha}}\left(\frac{2\mu_n}{\lambda-2\alpha}e^{\frac{\lambda-2\alpha}{2}(\phi-\phi_0)}\right)    +\mathsf{c}_+J_{\frac{\Tilde{\lambda}}{\lambda-2\alpha}}\left(\frac{2\mu_n}{\lambda-2\alpha}e^{\frac{\lambda-2\alpha}{2}(\phi-\phi_0)}\right)\right]\;,
\end{equation}
    with $\mathsf{c}_-$ and $\mathsf{c}_+$ constants depending on the initial conditions. It is easy to see that depending on the sign of $\lambda-2\alpha$ different behaviours occur:
    \begin{itemize}
        \item For $\alpha<\frac{\lambda}{2}$ the tower modes asymptotically decay as
        \begin{equation}
            \psi_n(\phi)\sim \exp\left\{\frac{2\alpha-\lambda-2\tilde\lambda}{4} \phi\right\} \cos\left[\frac{2\mu_n}{\lambda-2\alpha}\phi\right]\to 0\;,
        \end{equation}
        with their v.e.v. quickly decaying to 0.
        \item For $\alpha=\frac{\lambda}{2}$, the scalars decay as
        \begin{align}
            \psi_n(\phi)&= \mathsf{c}_- e^{\frac{1}{2}\left(-\Tilde{\lambda}-\sqrt{\Tilde{\lambda}^2-4\mu_n^2}\right)\phi}   +\mathsf{c}_+ e^{\frac{1}{2}\left(-\Tilde{\lambda}+\sqrt{\Tilde{\lambda}^2-4\mu_n^2}\right)\phi}\;.
        \end{align}
        Depending on whether
        \begin{equation}
            f(n)\lessgtr \frac{(d-2)\lambda^2}{8}\left[\left(\frac{2}{\lambda}\sqrt{\frac{d-1}{d-2}}\right)^2-1\right]\frac{H_0}{m_0}
        \end{equation}
        the amplitude will decay monotonously or oscillate. As we expect $f(n)$ to increase, there exists a $n_0\in\mathbb{N}$ such that all states $\psi_n$ with $n\geq n_0$ decay in an oscillating way. 
        \item For $\alpha>\frac{\lambda}{2}$ one finds that the two modes behave quite differently, with both a blowing-up mode and one for which asymptotically the states from the tower remain constant:
        \begin{subequations}
        \begin{align}
        \psi_n^{(-)}(\phi)&\sim e^{(2\alpha-\lambda)\phi}\to\infty\\
            \psi_n^{(+)}(\phi)&\sim \underbrace{\frac{\mathsf{c_{+}}}{\Gamma\left(1+\frac{\tilde\lambda}{\lambda-2\alpha}\right)}\left(\frac{2\alpha-\lambda}{\mu_n}\right)^{\frac{\tilde\lambda}{\lambda-2\alpha}}}_{\psi_n^{(\infty)}}+\mathcal{O}\left(e^{(\lambda-2\alpha)\phi}\right)\to \psi_n^{(\infty)}
        \end{align}
        \end{subequations}
        The blowing up mode $\psi_n^{(-)}(\phi)$ is not actually a solution, as replacing these solutions on the kinematical variables \eqref{e.kin var} results in the point $\Vec{X}_0=(0,0,0,1)$, which is not a fixed point of \eqref{eqs.dyn sis}. We are the left with the $\psi_n^{(+)}(\phi)$ mode, which asymptotes to a constant value $\psi_n^{(\infty)}$ which is always non-zero except for $\frac{\tilde\lambda}{\lambda-2\alpha}\in\mathbb{Z}_{<0}$.
    \end{itemize}
    \item \textbf{Modulus kination} \cite{Joyce:1996cp,Gouttenoire:2021jhk}: In this regime, the scalar potential and energy density of the tower rapidly decays (we assume that $\alpha<\sqrt{\frac{d-1}{d-2}}$, otherwise resulting in $K_\theta$), with the Universe expansion being driven by the kinetic energy of the modulus $\phi$, which together with the Hubble parameter scale as
    \begin{subequations}
        \begin{align}
            H&=\frac{H_0}{1+H_0(d-1)(t-t_0)}\\
            \phi&=\phi_0+\sqrt{\frac{d-2}{d-1}}\log\left[1+H_0(d-1)(t-t_0)\right]\;.
        \end{align}
    \end{subequations}
    Analogously as in the tracker solution, the dynamics \eqref{e:psin} of the tower are given by
    \begin{equation}
        \ddot\psi_n+\frac{(d-1)H_0}{1+(d-1) H_0(t-t_0)}\dot\psi_n+m_0^2 f(n)^2\left[1+H_0(d-1)(t-t_0)\right]^{-2\alpha\sqrt{\frac{d-2}{d-1}}}\psi_n=0\;,
    \end{equation}
    which can be rewritten as
    \begin{equation}
        \psi_n''+\tilde\mu_n^2 e^{2(\sqrt{\frac{d-2}{d-1}}-\alpha)\phi}\psi_n=0\,,\quad \text{with }\mu_n=\frac{m_0f(n)}{\sqrt{(d-1)(d-2)}H_0}\;,
    \end{equation}
    with straightforward solution
    \begin{equation}
        \psi_n(\phi)=\mathsf{c}_JJ_0\left(\frac{\tilde\mu_n}{\sqrt{\frac{d-2}{d-1}}-\alpha}e^{(\sqrt{\frac{d-2}{d-1}}-\alpha)\phi}\right)+\mathsf{c}_YY_0\left(\frac{\tilde\mu_n}{\sqrt{\frac{d-2}{d-1}}-\alpha}e^{(\sqrt{\frac{d-2}{d-1}}-\alpha)\phi}\right)\;,
    \end{equation}
    with $\mathsf{c}_J$ and $\mathsf{c}_Y$ constants set by initial conditions. Considering $\sqrt{\frac{d-2}{d-1}}>\alpha$, the solutions quickly decay as
    \begin{equation}
        \psi_n\sim e^{\frac{1}{2}(\sqrt{\frac{d-2}{d-1}}-\alpha)\phi}\cos\left(\frac{\tilde\mu_n}{\sqrt{\frac{d-2}{d-1}}-\alpha}e^{(\sqrt{\frac{d-2}{d-1}}-\alpha)\phi}\right)\to 0\;,
    \end{equation}
    while in the $\alpha=\sqrt{\frac{d-1}{d-2}}$ one has
    \begin{equation}
        \psi_n(\phi)=\mathsf{c}_1\cos\left(\tilde\mu_n\phi\right)+\mathsf{c}_2\sin\left(\tilde\mu_n\phi\right)\;,
    \end{equation}
    with $\mathsf{c_1}$ and $\mathsf{c_2}$ depending on initial conditions, and the amplitude of the scalars remains finite while oscillating.
    
    \item \textbf{Tower kination}: Finally, take the attractor $K_\theta$ for $\cos \theta\in\left(\max\{\alpha^{-1},\frac{2}{\lambda}\}\times\sqrt{\frac{d-1}{d-2}},1\right)$, with the value of $\theta$ indicating the proportion of the total kinetic energy carried by the modulus or the tower, set by the initial conditions. For this attractor to exist $\alpha>\sqrt{\frac{d-1}{d-2}}$ is  required. The Hubble scale and modulus evolve as
    \begin{subequations}
        \begin{align}
            H&=\frac{H_0}{1+H_0(d-1)(t-t_0)}\\
            \phi&=\phi_0+\sqrt{\frac{d-2}{d-1}}\cos\theta\log\left[1+H_0(d-1)(t-t_0)\right]\;,\label{e.tow kin 0}
        \end{align}
    \end{subequations}
    whilst the asymptotic evolution of the tower modes at the attractor point is set by
    \begin{equation}
        \ddot\psi_n+(d-1)H\dot\psi_n=0\;,
    \end{equation}
    with solution
     \begin{equation}\label{e.tow kin 1}
         \psi_n=\frac{\dot\psi_{n,0}}{(d-1)H_0}\log\left[1+H_0(d-1)(t-t_0)\right]\;,
     \end{equation} 
     with $\{\dot\psi_{n,0}\}_n$ such that
     \begin{equation} \label{e.tow kin 2}       \sin\theta=\dfrac{\kappa_d\sqrt{\sum_n\dot\psi_n^2}}{\sqrt{(d-1)(d-2)}H_0}\;.
     \end{equation}
     The actual value of $\theta_0$ associated with the tower kination attractor $K_{\theta_0}$ depends on the initial conditions, see Figure \ref{f.din2}.
     Note that this means that the v.e.v. of the scalars of the tower blows up and grows linearly with $\phi$, see Figure \ref{f.sol4} (see the $t$ scale is logarithmic, as is the evolution of $\phi$). From \eqref{e:psin} one can show that all scalars of the tower eventually pick up some velocity even if they were initially frozen.
\end{itemize}
The different solutions described above are depicted in Figure \ref{f.sols} for some initial states of the tower in $d=4$.

    \begin{figure}[h]
\begin{center}
\begin{subfigure}[b]{0.49\textwidth}
\center
\includegraphics[width=\textwidth]{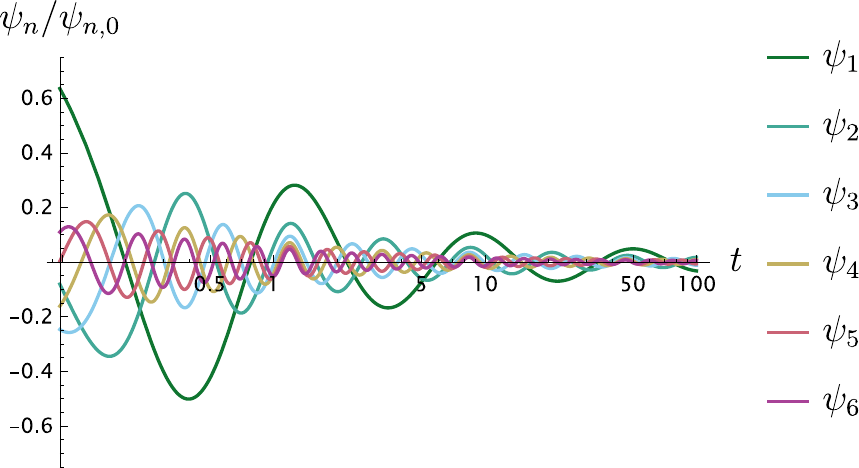}
\caption{\hspace{-0.3em}) $\lambda=\frac{3}{2}\sqrt{\frac{3}{2}}$, $\alpha=\sqrt{\frac{2}{3}}$. Decaying tower in \newline tracker solution.} \label{f.sol1}
\end{subfigure}\begin{subfigure}[b]{0.45\textwidth}
\center
\includegraphics[width=\textwidth]{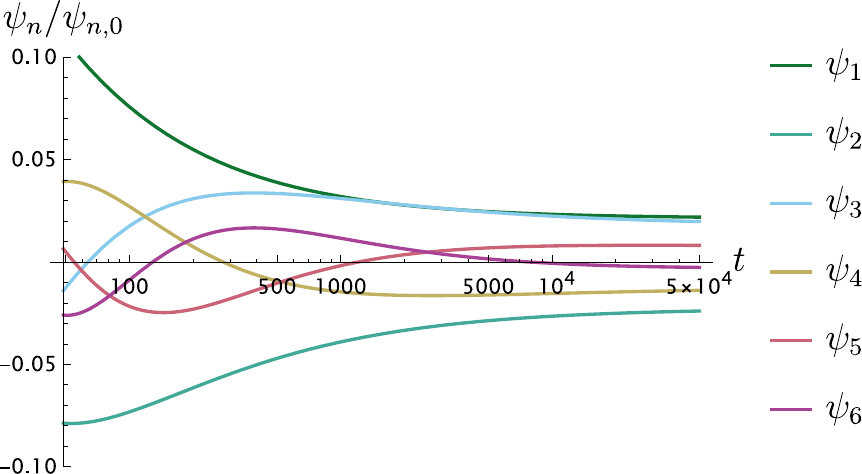}
\caption{\hspace{-0.32em}) 
 $\lambda=\frac{3}{2}\sqrt{\frac{3}{2}}$, $\alpha=\sqrt{\frac{3}{2}}$. Frozen tower in tracker solution.} \label{f.sol2}
\end{subfigure}
\hfill
\begin{subfigure}[b]{0.45\textwidth}
\center
\includegraphics[width=\textwidth]{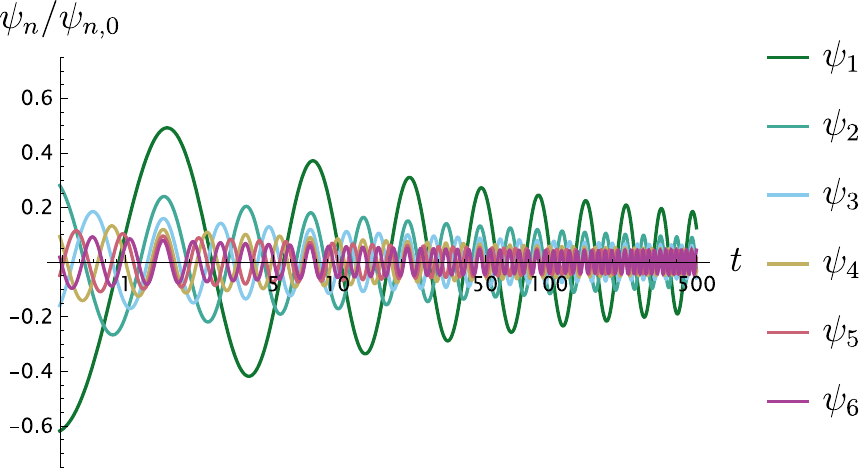}
\caption{\hspace{-0.3em}) 
 $\lambda=3\sqrt{\frac{3}{2}}$, $\alpha=\sqrt{\frac{2}{3}}$. Decaying tower\newline in modulus kination.} \label{f.sol3}
\end{subfigure}\begin{subfigure}[b]{0.45\textwidth}
\center
\includegraphics[width=\textwidth]{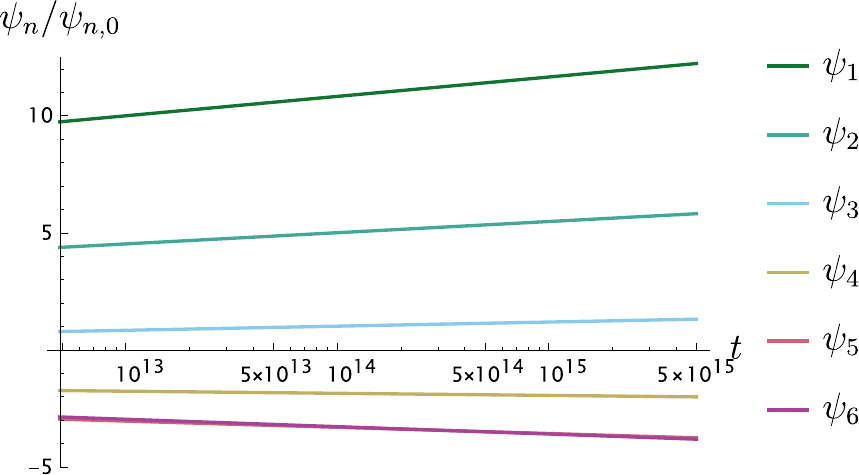}
\caption{\hspace{-0.3em}) $\lambda=2\sqrt{\frac{3}{2}}$, $\alpha=\sqrt{\frac{5}{2}}$. Growing scalars in tower kination.} \label{f.sol4}
\end{subfigure}
\hfill
\begin{subfigure}[b]{0.45\textwidth}
\center
\includegraphics[width=\textwidth]{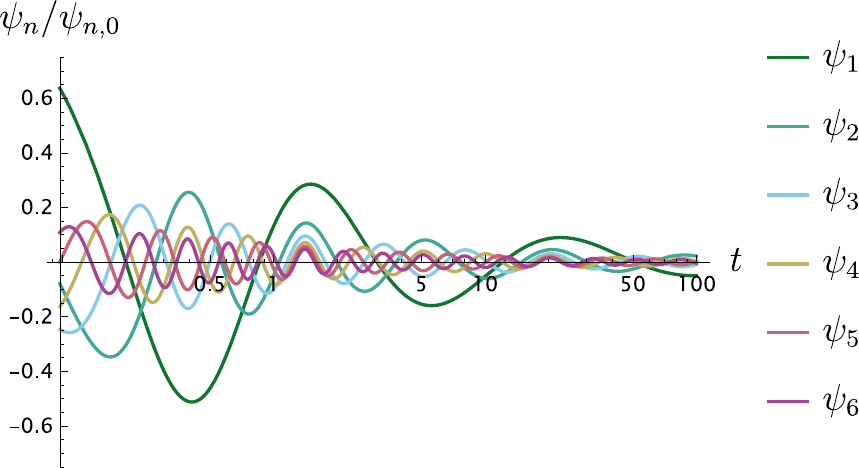}
\caption{\hspace{-0.3em}) $\lambda=\frac{3}{2}\sqrt{\frac{3}{2}}$, $\alpha=\frac{3}{4}\sqrt{\frac{3}{2}}$. Marginally \newline decaying scalars ($\lambda=2\alpha$) in tracker\newline solution.} \label{f.sol5}
\end{subfigure}\begin{subfigure}[b]{0.45\textwidth}
\center
\includegraphics[width=\textwidth]{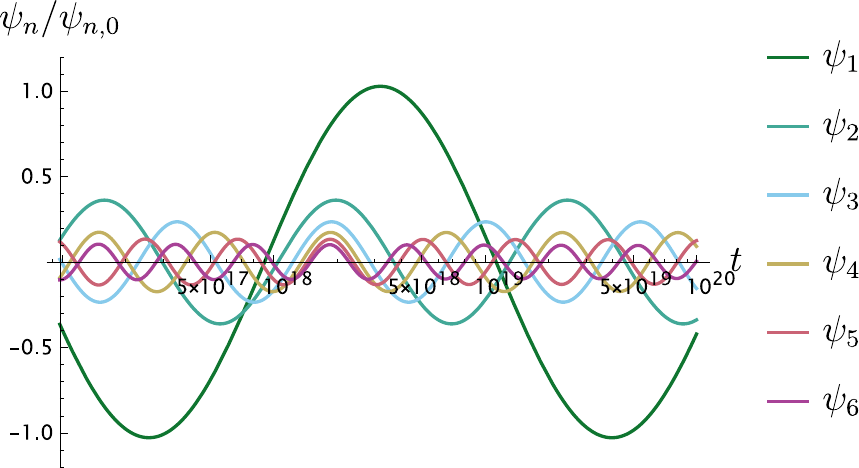}
\caption{\hspace{-0.3em}) $\lambda=3\sqrt{\frac{3}{2}}$, $\alpha=\sqrt{\frac{3}{2}}$. Constant amplitude, oscillating scalars in modulus kination.} \label{f.sol6}
\end{subfigure}
\caption{Evolution of the six first states $\{\psi_n\}_{n=1}^{6}$ of a tower with $f(n)=n$, as function of time $t$ (in arbitrary units and with interval values optimised for representation) for $d=4$. The initial values are taken $\psi_{n,0}\sim \psi_{1,0}n^{-1}$. The different regimes described in section \ref{sec:sols} are depicted.
\label{f.sols}}
\end{center}
\end{figure}

As a final comment, notice that in this attractor solution, there is a one-to-one correspondence between the cosmological time and the distance traversed in moduli space, with all numerical factors determined by the attractor point. In line with the \emph{species thermodynamics} proposal \cite{Cribiori:2023ffn}, there is a natural identification of time and moduli simply stemming from the existence of a light tower of states or a runaway potential. From a dynamical cobordism \cite{Buratti:2021yia,Buratti:2021fiv} interpretation, we can understand these solutions as bordisms between the initial $d$-dimensional EFT and the resulting $(D=d+n)$-dimensional theory after decompactification, with the UV effects being patent once the species scale $\Lambda_{\rm QG}\sim M_{{\rm Pl},D}$ is of the order of the EFT cut-off. Using the usual description in terms of \emph{critical exponent} $\delta$ \cite{Angius:2022aeq}, we can relate the spacetime $\Delta x$ and field space $\Delta \phi$ distances as
\begin{equation}\label{e.dyncob}
    \Delta \phi\sim \mp\frac{2}{\delta}\log \Delta x\;,
\end{equation}
where the overall sign depends on whether the spacetime distance is spacelike or timelike. As in our case we identify $\Delta x\equiv t$, we will consider $\Delta \phi\sim \frac{2}{\delta}\log \Delta x$.
Simple inspection of \eqref{e.at phi} allows us to identify the critical exponent in an attractor point $\vec X_0$ as
\begin{equation}
    \delta=2\frac{x_0^2+y_0^2}{x_0}\sqrt{\frac{d-1}{d-2}}=\left\{
    \begin{array}{ll}
         \lambda\in\left[\frac{2}{\sqrt{d-2}},2\sqrt{\frac{d-1}{d-2}},\right) &\;\text{for } T_\lambda\\
         2\sqrt{\frac{d-1}{d-2}} &\;\text{for } K_+\\
         \frac{2}{\cos\theta} \sqrt{\frac{d-1}{d-2}}>2\sqrt{\frac{d-1}{d-2}} &\;\text{for } K_\theta
    \end{array}
    \right.\;,
\end{equation}
so that in general $\delta\geq \frac{2}{\sqrt{d-2}}$ for the studied solutions. Note that while it is expected $\delta\leq 2\sqrt{\frac{d-1}{d-2}}$ for timelike dynamical cobordisms with positive potential \cite{Angius:2022mgh,Angius:2024zjv}, we find that for tower kination solutions $K_\theta$ this is always violated. One possible way of reconciliation is by considering that, as for these kind of solutions the tower scalars become light \emph{very fast} and are effectively massless (at least low-laying enough modes), they can be considered to contribute to the field space distance $\Delta \phi$. Simple inspection from the EFT action \eqref{e. EFT action}  and solutions \eqref{e.tow kin 0}, \eqref{e.tow kin 1} and \eqref{e.tow kin 2} result in generalised field space distance
\begin{equation}    \Delta\tilde\phi=\int_{\tau_0}^{\tau_f}\dd\tau\sqrt{\mathsf{G}_{ij}\partial_\tau\varphi^i\partial_\tau\varphi^j+\sum_n(\partial_\tau\psi_n)^2}
\end{equation}
obeying \eqref{e.dyncob} with $\delta= 2\sqrt{\frac{d-1}{d-2}}$ when the whole tower $\{\psi_n\}_n$ (not simply the low-laying modes) is taken into account. While this goes in the line of \cite{Mohseni:2024njl} using the Euclidean action as a proxy for distance between different theories, it has the potential problem of having to consider an infinite number of almost-massless states in our field distance.

\section{Upper bounds on exponential rates}\label{sec:dis}

In section \ref{sec:sols} we have shown how, depending on the values of the exponential factors $\alpha$ and $\lambda$, different asymptotic behaviours are featured for the light tower of states $\{\psi_n\}_n$. Among the different $(\lambda,\alpha)$ combinations, only three asymptotic behaviours are possible: \textbf{(a)} Tracker solution (both with frozen or decaying tower states), \textbf{(b)} modulus kination and \textbf{(c)} tower kination.
As shown in Figure \ref{f.exp_space}, these can be mapped to either decaying\footnote{For $\lambda>2\sqrt{\frac{d-1}{d-2}}$ and $\alpha=\sqrt{\frac{d-1}{d-2}}$ the amplitude of the scalars does not decay, but unlike the frozen or blowing-up behaviours, it oscillates.}, frozen or blowing-up scalars, depending on the asymptotic behaviour of the tower. Interestingly, decaying scalars always feature $\alpha\leq\min\left\{\frac{\lambda}{2},\sqrt{\frac{d-1}{d-2}}\right\}$.
This values for $\alpha$ were already conjectured in \cite{Rudelius:2022gbz} to be always the case, based on consistency under dimensional reduction and string theory examples. The other two possibilities, namely the tower kination $K_\theta$ ($\alpha>\sqrt{\tfrac{d-1}{d-2}}$ and $\lambda>2\sqrt{\tfrac{d-1}{d-2}}$) and the frozen tower $T_\lambda$ with $\lambda< 2\alpha$, are not found in construction in literature, nor they appear in the explicit examples that will be tested in section \ref{sec:pop}. 
The natural question that arises from the evidence is then whether there is anything fundamentally wrong with this kind of values for the exponential rates?

\vspace{0.75em}
From the naive EFT point of view, the two following problems appear when towers of states become light \emph{too fast}:
\begin{itemize}
    \item \textbf{Higher order expansion}:
    Having a large suppression in the masses compared to the runaway potential comes at the price of the appearance of higher power terms. In our study, we have assumed that these terms are negligible by imposing our tree-level term to be the dominant, namely, 
    \be
     \frac{m_{\psi_n}^2}{2}\psi_n^2 > |c_m| \frac{\psi_n^{2m + 4}}{\Lambda^{2m}}, \quad \forall\, n,m = 1,2,\dots\;,
    \ee
    where $\Lambda$ is the cut-off of our effective action, which will always be lower or equal than $\Lambda_{\rm QG}$, and $c_m$ are the Wilson coefficients in the higher power expansion,  so that the tower dynamics are governed by the mass terms.
    However, this can be rewritten as follows,
    \be
|\psi_n| \lesssim \left(\frac{\Lambda^{2m}m_{\psi_n}^2}{|c_m|}\right)^{\frac{1}{2m+2}}, \quad \forall\, n,m = 1,2,\dots
    \ee
    If the scalars $\{\psi_n\}_n$
    as seen in section \ref{sec:sols} freeze or blow up, for late times the hierarchy between the tree level and the higher-order terms is no longer satisfied and the EFT description ceases to be valid unless all $c_m$ decay fast enough. However, note that if this is not the case for some term $k_0$, then the field picks up an effective mass
    \begin{equation}
        m_{\psi_n,{\rm eff}}^2=\frac{1}{2}\partial_{\psi_n}^2\mathcal{L}_{\psi_n}=m_{\psi_n}^2+\sum_{k=0}^\infty (2k+4)(2k+3)\frac{c_k}{\Lambda^{2k}}\psi_n^{2k+2}\sim \frac{c_{k_0}}{\Lambda^{2 k_0}}\psi_{n}^{2k_0+2}\gg m_{\psi_n}^2\;.
    \end{equation}
    Independent of whether the tower freezes or blows up (note that this occurs linearly, not exponentially, with $\phi$), the effective mass becomes light parametrically slower than the bare $m_\psi\sim M_{{\rm Pl,}d}\; e^{-\alpha\phi}$. One can thus recursively argue that in the presence of higher power terms that are not suppressed fast enough, the tower picks up an effective mass which becomes light within the allowed regime, i.e., $\alpha_{\rm eff}\leq \min\{\frac{\lambda}{2},\sqrt{\frac{d-1}{d-2}}\}$.

    A study on how fast these higher power terms must fall in the EFT expansion is beyond the scope of this paper. Higher curvature terms can be used as a proxy for the species scale \cite{vandeHeisteeg:2022btw,vandeHeisteeg:2023ubh,Castellano:2023aum,vandeHeisteeg:2023dlw}, which in turn condition the scaling of the associated Wilson coefficients, but this is less clear for scalar power terms.

    \textcolor{black}{Related to this, the presence of a flux potential results in a ``ground mass'' for the metric fields (and their KK partners),\footnote{We thank Fernando Marchesano for pointing this scenario to us.}
    \begin{equation}
        \frac{V_F}{M_{{\rm Pl},d}^d}\sim \left(\frac{\mathcal{V}_{n}}{\mathcal{V}_{n,0}}\right)^{-\frac{d}{d-2}}\int_{X_n}F_p\wedge\star F_p\supseteq m_{\psi,*}^2\psi^2+...\;,
    \end{equation}
    with the different KK modes spaced by multiples of the KK mass $m_{\psi,0}$. In the diluted flux approximation the above contribution $m_{\psi,*}$ is neglected with respect to $m_{\psi,0}$, but if $\alpha\geq \frac{\lambda}{2}$, then eventually the mass of the different KK modes is effectively set by $m_{\psi,*}$, resulting in $\alpha_{\rm eff}=\frac{\lambda}{2}<\alpha$. However, this does not immediately solve the problem of having the KK tower decaying faster than the potential, as one on principle could turn off the fluxes, so that the above ground mass is not present and the above mechanism does not occur, provided the exponential rate $\alpha$ is independent of the value of the fluxes.}

    \item \textbf{Hubble scale and Higuchi bound}: For the frozen tower $T_\lambda$ with $\lambda< 2\alpha$, it is observed that the Hubble scale is parametrically lower compared to the tower's scale, with exponential scaling $H\sim H_0e^{-\lambda_H \phi}$, where  $\lambda_H=\min\left\{\frac{\lambda}{2},\sqrt{\frac{d-1}{d-2}}\right\}$. This leads to extra dimensions expanding beyond the Hubble horizon, signalling a departure from a $d$-dimensional FLRW cosmology.\footnote{From the Emergent String Conjecture perspective, one should also consider the possibility that the tower corresponds to oscillator modes coming from a lightly coupled emergent string. For the case $\lambda<2\alpha$ one would expect the local EFT to break down from the inclusion of higher spin modes. Moreover, from the universal scaling of the critical string mass, we expect in these cases $\alpha=\frac{1}{\sqrt{d-2}}$, for which $\alpha> \frac{\lambda}{2}$ results in a potential going against the Strong de Sitter conjecture.} For the case of tower kination $K_\theta$ $\big(\alpha>\sqrt{\frac{d-1}{d-2}}$ and $\lambda>2\sqrt{\frac{d-1}{d-2}}\big)$ the dynamics are characterised by all the scalars of the tower diverging. Assuming that the tower states presumably correspond to KK copies of a higher dimensional field, (e.g. Fourier modes of the radion), as we decompactify and make the internal dimensions larger, they would become more ``wrinkled'', eventually reaching curvatures in the decompactifying manifold of the order of the species scale.  An important point to make is that there are no examples in the literature that feature these kinds of behaviour, and as such a top-down study is out of reach, allowing only for qualitative, bottom-up arguments. Similar arguments requiring the Hubble scale to decay lower than the tower one were already discussed in \cite{Rudelius:2022gbz}.

    As a final comment on what \emph{might go wrong} for these kinds of scalings, we note that for both problematic regions of $(\lambda,\alpha)$ values the Higuchi bound   
    \begin{equation}
        m^2_{(l)}\geq H^2(l-1)(d+l-4)\;, 
    \end{equation}
    is violated, where $l\geq 2$ is the spin of the particle. Even though for the analysis of this paper we have assumed for simplicity scalar fields, for which the above bound does not apply, in usual string compactifications all KK replicas (including the graviton, with $l=2$) scale with the same exponential rate. While the original argument \cite{Higuchi:1986py} applied to positive cosmological constant (this is, actual de Sitter minima rather than runaway potentials), different generalisations have been proposed in the literature for theories with varying potentials \cite{Fasiello:2013woa,Luben:2020wim, Grisa:2009yy}. From this point of view, our bottom-up approach can be seen as a \emph{naive} justification for the de Sitter Higuchi bound to also apply in the case of runaway solutions. 
\end{itemize}

The above proposed restrictions, together with the Sharpened Distance Conjecture \cite{Etheredge:2022opl} and the Strong de Sitter bound \cite{Rudelius:2021azq} tightly constrain the possible values taken by $(\alpha,\lambda)$ to the blue region shown in Figure \ref{f.exp_space}. We could wonder whether there are further constraints from imposing consistency conditions. One can impose that the species scale $\Lambda_{\rm QG}$ does not become light faster than the potential $V$ and the mass of the leading tower, $m$, i.e.,
\begin{equation}\label{e.extra}
    \kappa_d\sqrt{V}\lesssim H\lesssim m\lesssim\Lambda_{\rm QG}\;.
\end{equation}
Assuming for simplicity that all $N$ states contributing to $\Lambda_{\rm QG}$ come from the leading tower (see \cite{Castellano:2023stg,Castellano:2023jjt} for general considerations when several towers are taken into account), with a density parameter $p$ \cite{Castellano:2021mmx} \footnote{For the case of KK towers, $p=n$ corresponds to $n$ dimensions becoming large while for the string tower it takes the value $p=\infty$. There are also more exotic states that come from open strings with $p=2$ \cite{Casas:2024ttx,Casas:2024clw}.} so that $m_n = n^{1/p}m_0 e^{-\alpha\phi}$,
one has
\be
\Lambda_{\rm QG}=m_N = N^{1/p} m_0 e^{-\alpha\phi}\Longrightarrow
\frac{\Lambda_{\rm QG}}{M_{{\rm Pl},d}}\sim e^{- \frac{p\,\alpha}{d-2 + p}\phi}.
\ee
From which one obtains
\be
\frac{p}{d-2+p}\leq1,\qquad \alpha \leq \frac{d-2+p}{2p}\lambda 
\ee
The first inequality is immediately fulfilled for $d\geq 2$, while the second one is trivialised for $\alpha\leq\frac{\lambda}{2}$ and saturated for $p=\infty$, the latter corresponding to the string oscillator spectrum. It is then clear that  \eqref{e.extra} does not impose further constraints on the possible values taken by $(\alpha,\lambda)$.

In \cite{Castellano:2023stg,Castellano:2023jjt} the following pattern was observed to hold in all examples from the literature (including multi-moduli ones):
\begin{equation}
    \frac{\vec\nabla m}{m}\cdot\frac{\vec\nabla \Lambda_{\rm QG}}{\Lambda_{\rm QG}}=\frac{1}{d-2}\;,
\end{equation}
which implies that the exponential factor at which $\Lambda_{\rm QG}$ scales fulfils $\mu\geq \frac{1}{(d-2)\alpha}$. This in turn results in
\begin{equation}
    \alpha\geq\frac{1}{\sqrt{d-2}}\;,\qquad \lambda\geq \frac{2}{\alpha(d-2)}\;,
\end{equation}
both of which are implied by the Sharpened Distance Conjecture and Strong de Sitter Conjecture.

Finally, one can wonder whether one can have a value of $\lambda$ arbitrarily large, i.e., whether a positive exponential potential can decay arbitrarily fast. One could argue against this in the case of positive potentials, which always results in SUSY being broken and an asymmetry between fermions and bosons. Depending on whether more light\footnote{Here \emph{light} refers in comparison to the KK scale.} degrees of freedom of one or other kind dominate, the associated Casimir potential $V_C$, resulting from 1-loop contributions \cite{Arkani-Hamed:2007ryu} to the energy momentum tensor, will be positive (more fermionic d.o.f.'s) or negative (more bosonic d.o.f.'s). Hence, scalar potential terms that become lighter than $V_C$ will be irrelevant, no matter the sign of $V_C$. This way, asymptotically \emph{positive} potentials, including also non-perturbative potentials, cannot decay faster than the Casimir energy, which in turn tells us that one always has an upper bound on the effective value of  $\lambda$. In scenarios where (e.g. non-perturbative) double-exponential potentials appear, one has that effectively $\lambda\equiv\infty$. When these are positive, SUSY is consequently broken, again resulting in the appearance of Casimir exponential runaway potentials.

The Casimir potential scales as $R_X^{-d}$ \cite{Arkani-Hamed:2007ryu} for large volumes, with $R_X\sim \mathcal{V}_X^{1/n}$ the characteristic length of the fastest growing internal $n$-cycle in higher-dimensional Planck units, so that
\begin{equation}
    \frac{V_C}{M_{{\rm Pl},d}^d}=\frac{V_C}{M_{{\rm Pl},D}^d}\left(\frac{M_{{\rm Pl},D}}{M_{{\rm Pl},d}}\right)\sim R_X^{-d\frac{d-2+n}{d-2}}\sim \mathcal{V}_X^{-d\frac{d-2+n}{n(d-2)}}\;.
\end{equation}
Assuming that this occurs homogeneously, the associated exponential rate is
\begin{equation}   
\lambda_C=\hat{\tau}\cdot\vec{\mu}_C\leq d\sqrt{\frac{d-2+n}{n(d-2)}}\leq d\sqrt{\frac{d-1}{d-2}}\;.
\end{equation}

On the other hand, the growth of an internal cycle also results in the tower of the associated KK modes becoming light. Again, for homogeneous decompactifications, $\alpha_{{\rm KK},n}=\hat{\tau}\cdot\vec{\zeta}_{{\rm KK},n}\leq\sqrt{\frac{d-2+n}{n(d-2)}}$. Being both Casimir and KK scales proportional, we find that $\vec{\mu}_C=d \vec{\zeta}_{{\rm KK},n}$, and as such $\lambda_C=d \alpha_{{\rm KK},n}$. Expecting the fastest decompactifying cycle to be associated with the lightest KK tower, one then reaches 
\begin{equation}
    (\lambda,\alpha)\in\left\{\frac{1}{\sqrt{d-2}}\leq \alpha\leq\sqrt{\frac{d-1}{d-2}}\;,2\alpha\leq\lambda\leq d\alpha\right\}\;,
\end{equation}
as depicted in the blue region in Figure \ref{f.puntos} for the case $d=4$.
Similar arguments on the relation between the scaling of the towers and the potential were already made in \cite{Rudelius:2021oaz,Gonzalo:2021fma,Castellano:2021mmx,Montero:2022prj}.

While the above discussion has assumed homogeneous decompactification of the internal volume, it is expected that the proposed bounds hold also for warped compactifications, as these seem to always result in lower exponential rates \cite{Etheredge:2022opl,RuizWARPED} while still fulfilling the previously commented bounds.

In summary, it is remarkable that simply from the study of the tower dynamics from the EFT perspective, some bounds for the exponential rate found in the literature can be replicated. In order to avoid  frozen scalars, the SDC automatically implies de $\lambda\geq \frac{2}{\sqrt{d-2}}$ and $\lambda\geq 2\alpha$ bounds, while the tower kination behaviour seems to indicate what goes wrong when the $\alpha
\leq\sqrt{\frac{d-1}{d-2}}$ bound is violated. On the other hand, from the naive EFT perspective nothing seems to go wrong for arbitrarily fast runaway exponential potentials, and extra arguments (i.e., the appearance of Casimir energies) must be made in order to find bounds in this direction.

\section{Populating the $(\alpha,\lambda)$ landscape}\label{sec:pop}
\begin{figure}
    \centering
    \includegraphics[width=1\linewidth]{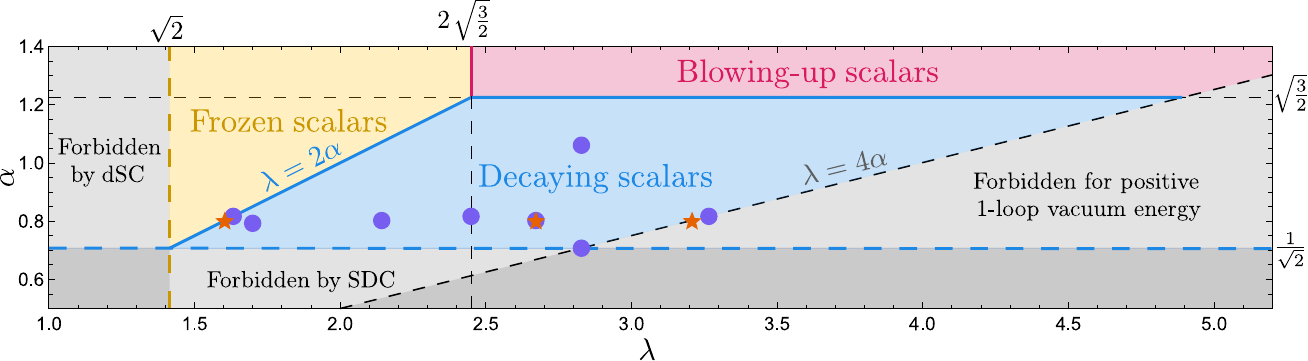}
    \caption{Values of the exponential factors $(\lambda,\alpha)$ from the M-theory compactifications in Section \ref{s.Mth} (orange stars) and the different flow trajectories in Type II compactifications shown in Figure \ref{f.dots} and Table \ref{t.flows} (lilac dots), with $d=4$. Note that all points correspond to the blue region (including its boundary), corresponding to $T_\lambda$ (with $\alpha\leq\frac{\lambda}{2}$) and $K_+$ attractors, see Figure \ref{f.exp_space}. For reference, the lines $\lambda=2\alpha,\,4\alpha$ have been pictured, with all the points lying between them. The $G7$ and $F6$ flux potentials in M-theory and IIA compactifications have not been pictured as they are never the leading term in the scalar potential.}
    \label{f.puntos}
\end{figure}

In this section, we will look at the different exponential rates $\alpha$ and $\lambda$ that appear in the different setups already considered in the literature. We will also look at their position in the study carried out in the previous section, and we will summarise the results graphically in Figure \ref{f.puntos}. For this purpose, we consider all the different terms that can appear in the scalar potential, e.g. from towers of states, fluxes, Casimir energies, curvature terms, etc.

\subsection{Compactifications of M-theory}\label{s.Mth}
For our first set of examples, we will consider the simple case of compactifying 11d M-theory on a $n$-dimensional compact manifold $X_n$, with volume $\mathcal{V}_n$ in 11d Planck units, down to $d=11-n$. For the sake of the approach, we will only consider the volume scaling, ignoring the rest of the possible moduli. A simple exercise of dimensional reduction results in a canonically normalised radion
 \begin{equation}
     \rho=\frac{3}{\sqrt{n(d-2)}}\log\frac{\mathcal{V}_n}{\mathcal{V}_{n,0}}\;,
 \end{equation}
parameterizing the light KK tower
\begin{equation}
    \frac{m_{\rm KK}}{M_{{\rm Pl},d}}   \sim e^{-\frac{3}{\sqrt{n(d-2)}}\rho}\;,
\end{equation}
so that in this case $\alpha=\frac{3}{\sqrt{n(d-2)}}=\frac{3}{\sqrt{(11-d)(d-2)}}$. Other possible towers of states, such as branes wrapped on internal cycles, will never become light faster than the associated species scale $\Lambda_{\rm QG}=M_{\rm Pl,11}\sim M_{\rm Pl,d} e^{-\frac{1}{3}\sqrt{\frac{n}{d-2}}\rho}$, and thus can be safely ignored. As for the potential, we find the following possibilities:
\begin{itemize}
    \item \textbf{Flux Potential}: If some internal $p$-cycle $\mathcal{C}_p$ (with $p\leq n$) is threaded by some units of $G4$ or $G7$ flux, this results in a positive flux potential in the lower-dimensional EFT:
    
    \begin{equation}
        \frac{V_{Gp}}{M_{{\rm Pl},d}^d}=\frac{1}{2}\left(\frac{\mathcal{V}_{n}}{\mathcal{V}_{n,0}}\right)^{-\frac{d}{d-2}}\int_{X_n}G_p\wedge\star G_p\sim\left\{
        \begin{array}{ll}
            0 & \text{if }n<p\\
            e^{-\frac{2}{3}\frac{p(d-2)+n}{\sqrt{n(d-2)}}\rho}&\text{if }n\geq p
        \end{array}
        \right.\;,
    \end{equation}
  
   This way,
   \begin{equation}
       \lambda_4=\left\{
       \begin{array}{ll}
            0&\text{for }d\geq 8\\
            \frac{2d+2}{\sqrt{(11-d)(d-2)}}&\text{for }7\geq d\geq 3
       \end{array}\right.\;,\qquad \lambda_7=\left\{
       \begin{array}{ll}
           0&\text{for }d\geq 5\\
             \frac{4d-2}{\sqrt{(11-d)(d-2)}}&\text{for }d= 4,\,3 
       \end{array}\right.
   \end{equation}
    When existing, all these potentials result in kination solutions for the scalars.
    \item \textbf{Curvature potential}: Compactifying on a negatively curved compact manifold results in a positive runaway potential in the lower dimensional manifold. For hyperbolic manifolds, the internal (negative) curvature is constant and all moduli but the overall volume are stabilised (see \cite{Kaloper:2000jb} and references therein), which motivates this kind of constructions for the examples of interest in this subsection. Through dimensional reduction one obtains
    \begin{equation}
        \frac{V_{\mathcal{R}}}{M_{{\rm Pl},d}^d}=-\frac{1}{2}\left(\frac{\mathcal{V}_{n}}{\mathcal{V}_{n,0}}\right)^{-\frac{d}{d-2}}\int_{X_n}\dd^n x\sqrt{g_n}\mathcal{R}_{g_n}\sim e^{-\frac{6}{\sqrt{n(d-2)}}\rho}\;,
    \end{equation}
    and as such $\lambda_{\mathcal{R}}=\frac{6}{\sqrt{n(d-2)}}=\frac{6}{\sqrt{(11-d)(d-2)}}$. Note that this precisely lies in the $\alpha=\frac{\lambda}{2}$ line for the tracker region.
    \item \textbf{Casimir Potential}: As explained in section \ref{sec:dis}, supersymmetry breaking results in an unbalance between the fermionic and bosonic degrees of freedom in our theory. When more light fermionic than bosonic degrees of freedom are present in our non-supersymmetric theory, this results in a positive Casimir potential which asymptotically scales as
    \begin{equation}
        \frac{V_{C}}{M_{{\rm Pl},d}^d}\sim e^{-\frac{3d}{\sqrt{n(d-2)}}\rho}\;,
    \end{equation}
    resulting in $\lambda_C=\frac{3d}{\sqrt{n(d-2)}}=\frac{3d}{\sqrt{(11-d)(d-2)}}$, with the point $(\lambda,\alpha)$ laying in the moduli kination region for all values of $d$.
\end{itemize}
As also argued in section \ref{sec:dis}, when breaking supersymmetry (this is always the case when a \emph{positive} potential is present) there is always a Casimir contribution to the scalar potential. In this case, this translates in $\lambda_C$, with value $\frac{3d}{\sqrt{(11-d)(d-2)}}$, serving as an upper bound on $\lambda$. This is precisely what occurs for the $G7$ fluxes in M-theory or $F6$ in IIA compactifications, which go to zero faster than the associated Casimir, and as such they can never be the leading term in the potential.

\subsection{4d compactifications of Type II string theory.}

We now consider a (slightly) more involved set of compactifications, corresponding to type II string theories on Calabi-Yau orientifolds to obtain 4d $\mathcal{N}=1$ theories. For simplicity the only moduli we will consider are the 10d dilaton $\phi$ and compact volume $\mathcal{V}$ in 10d Planck units, which can be canonically normalised to
\begin{equation}
    \hat{\phi}=\frac{1}{\sqrt{2}}\phi\,,\qquad\rho=\sqrt{\frac{2}{3}}\log\frac{\mathcal{V}}{\mathcal{V}_n}\;.
\end{equation}
In the type IIA compactifications, depending on the trajectory taken the leading towers can correspond to KK modes associated with the decompactification of the whole volume,  D0 branes resulting from decompactification to 5d (M-theory on the Calabi-Yau), or the string oscillator modes \cite{Corvilain:2018lgw,Lee:2019wij}:
\begin{equation}
    \frac{m_{\rm KK}}{M_{\rm Pl,4}}\sim e^{-\sqrt{\frac{2}{3}}\rho}\;,\quad\frac{m_{\rm D0}}{M_{\rm Pl,4}}\sim e^{-\frac{3}{2\sqrt{2}}\hat\phi-\frac{1}{2}\sqrt{\frac{3}{2}}\rho},\;\quad\frac{m_{\rm osc}}{M_{\rm Pl,4}}\sim e^{\frac{1}{2\sqrt{2}}\hat\phi-\frac{1}{2}\sqrt{\frac{3}{2}}\rho}\;.
\end{equation}
As for the IIB case, apart from the $m_{\rm KK}$ and $m_{\rm osc}$ towers, we can obtain also the KK tower associated to the dual Calabi-Yau under 6 T-dualities, as well as the oscillator modes from the string resulting from wrapping the D7-brane over the whole Calabi-Yau (seen as the S-dual D1-string in the T-dual 3-fold)\cite{Lanza:2021qsu}:
\begin{equation}
    \frac{m_{\rm KK-T}}{M_{\rm Pl,4}}\sim e^{\frac{1}{\sqrt{2}}\hat{\phi}-\frac{1}{\sqrt{6}}\rho}\,\quad \frac{m_{\rm D7}}{M_{\rm Pl,4}}\sim e^{\frac{1}{\sqrt{2}}\hat{\phi}}
\end{equation}
In order to better quantify the different limits, we can introduce the scalar charge-to-mass ratio vectors \cite{Calderon-Infante:2020dhm,Etheredge:2022opl,Etheredge:2023odp} of the towers,
\begin{equation}
    \zeta^a_I=-\delta^{ab}\mathsf{e}_b^i\partial_{\phi^i}\log \left(\frac{m_I}{M_{{\rm Pl,}\, d}}\right)\;,
\end{equation}
where $\mathsf{e}_b^i$ is an inverse vielbein associated with the moduli space metric, resulting in
\begin{equation}
\begin{array}{c}
     \vec\zeta_{\rm KK}=\left(0,\sqrt{\frac{2}{3}}\right)\;,\quad \vec\zeta_{\rm D0}=\left(\frac{3}{2\sqrt{2}},\frac{1}{2}\sqrt{\frac{3}{2}}\right)\;,\quad \vec\zeta_{\rm osc}=\left(-\frac{1}{2\sqrt{2}},\frac{1}{2}\sqrt{\frac{3}{2}}\right)\,,  \\
     \vec{\zeta}_{\rm KK'}=\left(-\frac{1}{\sqrt{2}},\frac{1}{\sqrt{6}}\right) \;,\quad \vec{\zeta}_{\rm D7}=\left(-\frac{1}{\sqrt{2}},0\right)
\end{array}
\end{equation}
The associated exponential rate will be given by $\alpha_I=\hat{\tau}\cdot\vec\zeta_I$, with $\hat{\tau}$ the normalised tangent vector to our trajectory.
 \begin{figure}
    \centering
    \begin{subfigure}[b]{0.49\textwidth}
\center
\includegraphics[width=0.95\textwidth]{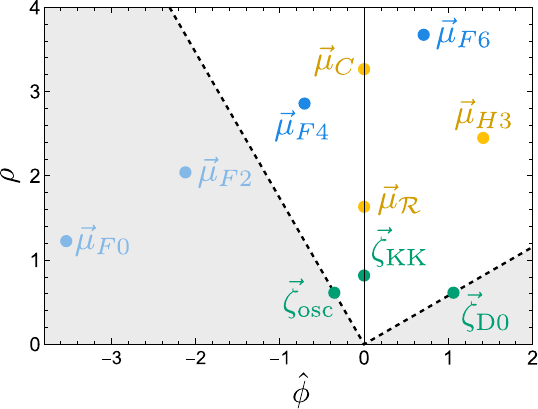}
\caption{\hspace{-0.3em}) Type IIA flux compactifications.} \label{f.dotsA}
\end{subfigure}
\begin{subfigure}[b]{0.49\textwidth}
\center
\includegraphics[width=0.95\textwidth]{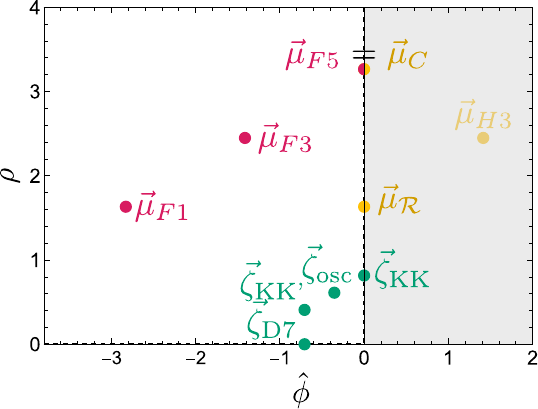}
\caption{\hspace{-0.3em}) Type IIB flux compactifications.} \label{f.dotsB}
\end{subfigure}
    \caption{Scalar charge to mass ratio vectors and de Sitter ratios for the different towers (green) and terms of the potentials considered for the Type IIA (blue), IIB (magenta) or both (yellow). In gray the regions of moduli space which do not correspond to this duality frame perturbative regime are portrayed.}
    \label{f.dots}
\end{figure}
As for the different potentials, we will consider different constructions. From the 10d Einstein frame action in type II supergravities \cite{Ibanez:2012zz,VanRiet:2023pnx},
\begin{equation}
    S_{\rm II}^{\rm (10 d)}\supseteq\frac{1}{2\kappa_{10}^2}\int\dd^{10} x\,\sqrt{-G}\left\{\mathcal{R}_G-\frac{1}{2\cdot 3!}e^{-\phi}H_3^2-\frac{1}{2}\sum_ne^{\frac{5-n}{2}\phi}\frac{F_n^2
    }{n!}\right\}\;,
\end{equation}
the following flux potentials in 4d can be obtained:
\begin{equation}\label{e. flux potential}
    \frac{V_F}{M_{\rm Pl,4}^4}\sim \mathsf{V}_{H3}e^{-\sqrt{2}\phi-\sqrt{6}\rho}+\sum_{n}\mathsf{V}_{Fn}e^{\frac{5-n}{\sqrt{2}}\hat{\phi}-\frac{n+3}{\sqrt{6}}\rho}\;,
\end{equation}
with $n=1,\,3,\,5$ for Type IIB and $n=0,\,2,\,4,\,6$ for type (possibly massive) IIA. For completeness sake, we can also consider the scaling of (positive) Casimir and curvature\footnote{While so far we have assumed a flat internal manifold, we could consider the potential term appearing from using a compact hyperbolic 6-manifold.} contributions to the potential:
\begin{equation}\label{e. extra potential}
    \frac{V_C}{M_{\rm Pl,4}^4}\sim e^{-4\sqrt{\frac{2}{3}}\rho}\;,\qquad\frac{V_{\mathcal{R}}}{M_{\rm Pl,4}^4}\sim e^{-2\sqrt{\frac{2}{3}}\rho}\;.
\end{equation}

\begin{table}[h!]
    \centering    
   \begin{tabular}{|c|c|c|c|}
    \hline
    \rowcolor{gray!10!}
       \centering Potential  & $\hat\tau$ & $\lambda$ & $\alpha$\\
       \hline
        $\mathcal{R},\,(H3,\,{\color{our-magenta}F1},\,{\color{our-magenta}F3},\,{\color{our-blue}F4},\,{\color{our-magenta}F5}/C,\,{\color{our-blue}F6})$&$\left(0,1\right)$&$2\sqrt{\frac{2}{3}}$&$\sqrt{\frac{2}{3}}$\\
        $H3,\,{\color{our-magenta}F3},\,({\color{our-magenta}F1},\,{\color{our-magenta}F5}/C)$&$\left(0,1\right)$&$\sqrt{6}$&$\sqrt{\frac{2}{3}}$\\
        $H3,\,{\color{our-blue}F4},\,(C,\,{\color{our-blue}F6})$&$\left(\frac{1}{2\sqrt{7}},\frac{3}{2}\sqrt{\frac37}\right)$&$5\sqrt{\frac{2}{7}}$&$\frac{3}{\sqrt{14}}$\\
        $H3,\,({\color{our-magenta}F5}/C,\,{\color{our-blue}F6})$&$\left(\frac{1}{2},\frac{\sqrt{3}}{2}\right)$&$2\sqrt{2}$&$\frac{3}{2\sqrt{2}}$\\
        ${\color{our-magenta}F3},\,({\color{our-magenta}F1},\,{\color{our-magenta}F5}/C)$&$\left(-\frac{1}{2},\frac{\sqrt{3}}{2}\right)$&$2\sqrt{2}$&$\frac{1}{\sqrt{2}}$\\
        ${\color{our-blue}F4},\,(C,\,{\color{our-blue}F6})$&$\left(-\frac{1}{2}\sqrt{\frac{3}{13}},\frac{7}{2\sqrt{13}}\right)$&$\sqrt{\frac{26}{3}}$&$\frac{7}{\sqrt{78}}$\\
        ${\color{our-magenta}F5}/C$&$\left(0,1\right)$&$4\sqrt{\frac{2}{3}}$&$\sqrt{\frac{2}{3}}$\\
        ${\color{our-magenta}F1},\,H3,\, ({\color{our-magenta}F3},\,{\color{our-magenta}F5}/C)$&$\left(-\frac{1}{2\sqrt{7}},\frac{3}{2}\sqrt{\frac{3}{7}}\right)$&$4\sqrt{\frac{2}{7}}$&$\frac{3}{\sqrt{14}}$\\
        ${\color{our-magenta}F1}$ & $\left(-\frac{\sqrt{3}}{2},\frac{1}{2}\right)$&$4\sqrt{\frac{2}{3}}$&$\sqrt{\frac{2}{3}}$\\
         \hline
    \end{tabular}
    \caption{Values of the exponential factors $(\lambda,\alpha)$, as well as the asymptotic direction $\hat{\tau}$, for different potentials in type {\color{our-blue}IIA} and {\color{our-magenta}IIB} compactifications. Note that the F5 and Casimir terms have the same scaling. The terms within the parenthesis are subleading and do not affect the asymptotic the $\lambda$ and $\alpha$ values.}
    \label{t.flows}
\end{table}
In the same way as for the towers, we can better study the moduli dependence of the different terms of the potential by introducing the so-called \emph{de Sitter ratios} \cite{Calderon-Infante:2022nxb}:
\begin{equation}
    \mu^a_J=-\delta^{ab}\mathsf{e}_b^i\partial_{\phi^i}\log \left(\frac{V_J}{M_{{\rm Pl},\, 4}^2}\right)\;,
\end{equation}
which are immediately read from our expressions \eqref{e. flux potential} and \eqref{e. extra potential} as they are given in canonically normalised moduli. In order to obtain the trajectories that our moduli will follow, we will consider the gradient flow of the scalar potential. The resulting direction can be obtained by considering the convex hull of the different non-zero de Sitter ratios $\{\vec\mu\}_J$ obtaining the point closest to the origin \cite{Calderon-Infante:2022nxb, Shiu:2023fhb}. The $\lambda$ of our trajectory will be simply given by the distance from the origin to the convex hull, and $\alpha$ by the projection of the leading scalar charge-to-mass ratio along the resulting direction $\hat{\tau}$. This is pictured in Figure \ref{f.dots}, with the different possibilities registered in Table \ref{t.flows} and Figure \ref{f.puntos}. In order to stay in the perturbative regime of large volume and small coupling \cite{Lanza:2021qsu}, we will require for type IIA compactifications
\begin{equation}\label{e.pert reg}
    \left.\begin{array}{rl}
         \mathcal{V}&\gg \ell_s^6  \\
         \varphi_4=\phi-\frac{1}{2}\log\frac{ \mathcal{V}}{\ell_s^6}&<0 
    \end{array}\right\}\Longrightarrow \rho\leq\max\left\{\frac{\phi}{\sqrt{3}},-\sqrt{3}\phi\right\}
    \;,
\end{equation}
where $\ell_s$ and $\varphi_4$ are the string length and the 4d dilaton. Note that the above region precisely corresponds with the directions spanned between $\vec\zeta_{\rm osc}$ and $\vec\zeta_{\rm D0}$. As depicted in Figure \ref{f.dotsA}, in order to stay in this region we do not consider directions following the gradient flows of the $F0$ (Romans mass), $F1$ and $F2$ terms of the potential. Furthermore, as was the case for $G7$ fluxes in M-theory compactifications in Section \ref{s.Mth}, the $F6$ term always decays faster than the (positive) Casimir term in the potential, and as such cannot dominate asymptotically. Because of this we do not consider it in Table \ref{t.flows} or Figure \ref{f.dots}. For the type IIB case, due to the T- and S-dualities, the K\"ahler cone is confined to the region $\hat{\phi}<0$ and $\rho>0$, spanned by the $\vec{\zeta}_{\rm KK}$ and $\vec{\zeta}_{\rm D7}$ directions. As evident from Figure \ref{f.dotsB}, the gradient flows associated to all RR flux potentials in type IIB compactifications keep us in this perturbative regime, whereas the $H3$ potentials leads us away from this region.

\section{Conclusions and outlook}\label{sec:conc}

Much has been learned in recent years about the constraints quantum gravity imposes on lower dimensional EFTs, especially in asymptotic regions of moduli space. However, it has not been until even more recently that the so-called \emph{Swampland program ``Precision Era''} has started, with considerable effort being put in the precise characterisation of the different $\mathcal{O}(1)$ factors involved in the different conjectures. In the present paper, we have used a bottom-up perspective to bound the possible exponential scalings of both runaway potentials and towers of light states, by analysing the dynamical cosmological evolution that EFTs with different parameters may have. Although these scenarios are not meant to be phenomenologically realistic, they show that for some values of the exponential factors  certain inconsistencies appear, which seem to suggest that they will not appear in EFTs with UV completion.

We have studied the evolution of FLRW cosmologies in the presence of an exponential runaway potential (with exponential rate $\lambda$) and a tower of light scalar states (with exponential rate $\alpha$). We have characterised the different behaviours in three types of solutions: \textbf{(a)} Tracker solution with $0\leq\lambda\leq 2\sqrt{\frac{d-1}{d-2}}$, where the scalar decay for $2\alpha\leq\lambda$ and remains frozen for $2\alpha>\lambda$, \textbf{(b)} modulus kination with $\lambda\geq 2\sqrt{\frac{d-1}{d-2}}$ and $\alpha\leq \sqrt{\frac{d-1}{d-2}}$, for which the tower quickly asymptotes to zero, and \textbf{(c)} tower kination with $\lambda\geq 2\sqrt{\frac{d-1}{d-2}}$ and $\alpha> \sqrt{\frac{d-1}{d-2}}$, where the tower is effectively massless and the v.e.v. of all its states diverges.

While naively all the above behaviours are solutions to the equations of motion, there are reasons to believe that some of them might be deeply flawed. The reason for this is for $\alpha>\min\left\{\frac{\lambda}{2},\sqrt{\frac{d-1}{d-2}}\right\}$ the effective theory ceases to be valid. This can be motivated by the lack of scale separation between the Hubble scale and the tower scale, with the internal dimensions growing parametrically faster than the external ones, as well as the takeover of higher correction terms over the tree level as the latter rapidly becomes smaller.

All these problems point to the conclusion that a possible UV embedding might not be feasible and, therefore that the light tower of states expected from the Swampland Distance Conjecture cannot decay arbitrarily fast, at least in constructions resulting in a FLRW cosmology. As a stimulating point, all these problematic solutions seem to violate the Higuchi bound for \emph{de Sitter space}, though it would be interesting to check whether this is modified for (probably too steep) \emph{runaway potentials}.

Additionally, in order to give evidence for these claims, in section \ref{sec:pop}, we confront them with a series points $(\lambda, \alpha)$ in the exponential factor space coming from different top-down M-theory and type II compactifications. This is only a first attempt in scratching the surface of $(\lambda, \alpha)$ values that can be obtained from top-down constructions, and a future systematic study in this direction would strengthen and narrow the rate at which the terms of our effective action can decay, which would prove extremely useful for phenomenology purposes, especially in cosmology.

\bigskip
{\bf Acknowledgments:}
We are grateful to Fernando Marchesano, Miguel Montero and Irene Valenzuela for inspirational conversations and comments on the manuscript. We would also like to thank Fien Apers, Matilda Delgado, Tom Rudelius, Bruno Valeixo Bento and Max Wiesner for illuminating discussions. I.R. wishes to acknowledge the hospitality of the Department of Physics of Harvard University and the Erwin Schr\"odinger International Institute for Mathematics and Physics of the University of Vienna during different stages of this work. G.F.C. and I.R. acknowledge the support of the Spanish Agencia Estatal de Investigaci\'on through the grant “IFT Centro de Excelencia Severo Ochoa” CEX2020-001007-S and the grant PID2021-123017NB-I00, funded by MCIN/AEI/10.13039/ 501100011033 and by ERDF A way of making Europe. G.F.C. is supported by the grant PRE2021-097279 funded by MCIN/AEI/ 10.13039/501100011033 and by ESF+. The work of I.R. is supported by the Spanish FPI grant No. PRE2020-094163 and ERC Starting Grant QGuide101042568 - StG 2021.

\vspace*{.5cm}
\bibliographystyle{JHEP2015}
\bibliography{ref}

\end{document}